\documentclass[aps,prd,showpacs,nofootinbib,floats,floatfix,preprintnumbers,groupedaddress,twocolumn]{revtex4-1}
\usepackage{graphicx,epsfig}
\usepackage{xcolor}
\usepackage{dcolumn}
\usepackage{bm}
\usepackage{latexsym}
\usepackage{booktabs}
\usepackage{color}
\usepackage{tabularx}
\usepackage{ulem}
\usepackage{hyperref}
\usepackage{float}
\usepackage{tabularx}
\usepackage{color}
\usepackage{comment}
\usepackage{physics}
\usepackage{subfigure}
\usepackage{tikz}
\usepackage{siunitx}
\usepackage{hyperref}
\usepackage{colortbl}
\usepackage{listings}
\usepackage{amsmath,amsfonts,amssymb}
\usepackage{fancyhdr}
\usepackage{orcidlink}
\usepackage{hyperref}
\usepackage{natbib}
\usepackage{multirow}
\usepackage{todonotes}
\usepackage{lipsum}

\hypersetup{
	colorlinks   = true, 
	urlcolor     = blue, 
	linkcolor    = blue, 
	citecolor    = blue    
}

\hypersetup{
	colorlinks   = true, 
	urlcolor     = blue, 
	linkcolor    = blue, 
	citecolor    = blue 
}
\def\o{\omega} 
\def\r{\rho}
\def\A{\mathcal{A}}
\def\no{\nonumber}

\def\O{\Omega}

\def\p{\partial}

\definecolor{lightblue}{RGB}{173,216,230}
\definecolor{lightgreen}{RGB}{144,238,144}
\definecolor{magenta}{RGB}{255,0,255}
\definecolor{myolive}{RGB}{128,128,0}
\definecolor{masteredyellow}{RGB}{255,255,102}
\definecolor{mymaroon}{RGB}{128,0,0}
\begin{document}
 
	\title{Dynamical analog spacetimes from nonlinear perturbations in a topological material}
	
        \author{Surajit Das$^{1,2}$\orcidlink{0000-0003-2994-6951}}
        \email{surajitdas@mail.ustc.edu.cn, surajit.cbpbu20@gmail.com}

	\author{Surojit Dalui$^{3}$\orcidlink{0000-0003-1003-8451}}
        \email{surojitdalui@shu.edu.cn}

        \author{Hrishit Banerjee$^{4,5}$\orcidlink{0000-0001-7852-497X}}
        \email{hbanerjee001@dundee.ac.uk}

        \author{Arpan Krishna Mitra$^{6,7}$\orcidlink{0000-0002-9865-252X}}
        \email{arpankmitra@gmail.com (Corresponding author)}

        \affiliation{
        $^{1}$Department of Astronomy, School of Physical Sciences, \textcolor{blue}{University of Science and Technology of China}, Hefei, Anhui 230026, China\\
        \\
        $^{2}$CAS Key Laboratory for Researches in Galaxies and Cosmology, School of Astronomy and Space Science, \textcolor{blue}{University of Science and Technology of China}, Hefei, Anhui 230026, China\\
        \\
        $^{3}$Department of Physics, \textcolor{blue}{Shanghai University}, 99 Shangda Road, Baoshan District, Shanghai 200444, China\\
        \\
        $^{4}$School of Science and Engineering, \textcolor{blue}{University of Dundee}, Dundee, Scotland, UK\\
        \\
        $^{5}$ School of Metallurgy and Materials, \textcolor{blue}{University of Birmingham}, Edgbaston, Birmingham, UK\\
        \\
        $^{6}$\textcolor{blue}{Aryabhatta Research Institute of Observational Sciences (ARIES)}, Manora Peak Nainital 263001, Uttarakhand, India\\
        \\
        $^{7}$Department of Physics, \textcolor{blue}{Harish-Chandra Research Institute}, Chhatnag Road, Jhunsi, Allahabad 211019, India
        }

\begin{abstract}
Emergent spacetime analogs in condensed matter systems have opened a fascinating window into simulating aspects of gravitational physics in controlled laboratory environments. In this work, we develop a comprehensive nonlinear analog gravity framework within a topological material, incorporating the impact of Berry curvature on the hydrodynamic flow of electrons. Unlike prevalent studies in existing literature limited to linear perturbations, we derive and analyze a fully nonlinear wave equation governing radial perturbations of density and velocity fields, which dynamically generate an effective acoustic metric. Taking the example of graphene as a representative system, and calculating its properties from first principles, we numerically demonstrate the formation of evolving acoustic horizons and quantify analog Hawking temperatures in experimentally accessible regimes. Our findings suggest that topological materials can serve as versatile platforms to probe rich gravitational phenomena, including horizon dynamics and quasi-thermal emission, beyond conventional linear approximations. This work lays the groundwork for exploring nonlinear emergent spacetime in a broad class of quantum materials, bridging condensed matter physics and gravitational analogs.
\end{abstract}

\maketitle

\section{Introduction}\label{sec:intro}

Analogies have long served as fertile ground for physical insight, linking disparate domains through shared mathematical structure. Within this spirit, the field of analog gravity explores condensed matter systems whose perturbative dynamics mimic those of fields in curved spacetime. This approach, first brought into sharp focus by Unruh’s seminal work \cite{Unruh:1980cg}, demonstrating that linear perturbations of velocity potentials in barotropic, inviscid, irrotational fluids obey a d’Alembertian equation on an emergent curved geometry, has since evolved into a rich and multidisciplinary program \cite{Euve:2015vml,Steinhauer:2015saa,MunozdeNova:2018fxv,Drori:2018ivu,Clovecko:2018qnj,Vocke:2018xwb,Solnyshkov:2018dgq,Aguero-Santacruz:2020krw,Petty1:2020,Leonhardt:2020fdi,Jacquet:2020znq,Blencowe:2020ygo,Bera:2020doh,Anacleto:2013esa,Jacquet:2020bar}. The acoustic metric derived in such settings plays the role of a surrogate gravitational background, giving rise to ``dumb holes"—analogs of black holes that trap phonons in a manner mathematically analogs to light trapped in a spacetime horizon \cite{Barcelo:2011fc}.

The traditional analog gravity paradigm predominantly relies on linearized perturbations about a fixed, stationary background. This simplification, while providing important theoretical and experimental pathways — including laboratory simulations of Hawking radiation \cite{Blencowe:2020ygo}, quasinormal modes \cite{Torres:2020tzs,Liu:2024vde}, and superradiance \cite{Patrick:2020baa} — remains intrinsically limited in its capacity to capture the full dynamical richness of emergent geometries. Indeed, general relativistic spacetimes of astrophysical interest are inherently dynamical, evolving under gravitational collapse, mergers, or accretion. Extending analog gravity beyond the linear regime to include nonlinear perturbations is thus a vital step toward simulating more realistic spacetime phenomena in the laboratory and deepening our understanding of emergent geometry in physical systems \cite{Sen:2012rb,Albuquerque:2025eny,MalatoCorrea:2025iuc}.

Recent work has begun to bridge this gap by Fernandes, Maity, and Das \cite{Fernandes:2021gkf,Fernandes:2022bwo}, in which they developed a formalism of the full nonlinear dynamics of inviscid, barotropic, spherically symmetric flows were re-expressed in terms of perturbations of the mass accretion rate, rather than the velocity potential. This shift enables the derivation of a nonlinear wave equation where all perturbative orders contribute to a dynamical analog metric. Unlike the traditional acoustic metric, which is fixed by background parameters, the new metric components evolve with the perturbations, admitting time dependent acoustic horizon that may grow or recede in response to perturbations. This framework marks a significant conceptual advancement, moving from kinematic analogies to fully dynamical analog spacetime.

Current developments in fluid dynamics have introduced geometric and topological corrections into hydrodynamic frameworks. In particular, Mitra et al. \cite{Mitra:2021wjp} demonstrated that Eulerian fluid variables can obey an anomalous dynamical algebra arising from modified Poisson brackets that incorporate Berry curvature corrections. These corrections originate from a quantum input — namely, the generalized bracket structure satisfied by discrete Lagrangian fluid elements — and closely mirror the semiclassical behavior of electrons in magnetic Bloch bands, where a periodic lattice potential, external magnetic field, and Berry curvature, all influence the motion \cite{Sundaram:1999zz,Xiao:2009rm,Nagaosa:2009ycg,Abir:1972,Haldane:2004zz,Nielsen:1983,Bohm:2003,Duval:2005vn,Gosselin:2006ht,Chang:1995zz}. Such corrections enrich the phase space structure and open the door to explore fluid behavior influenced by quantum geometry. Electron hydrodynamics, where charge transport is governed by collective fluid-like motion rather than conventional Ohmic behavior, offers a compelling condensed matter realization of these theoretical ideas \cite{ssg:1977,Slav:2010,Polini:2020}.

Remarkably, recent experiments have established the hydrodynamic regime in ultra-clean electronic systems such as graphene \cite{Lucas:2017idv} and high-conductivity layered materials like the metallic delafossites $PdCoO_2$ and $PtCoO_2$ \cite{Andy:2016}. These materials possess nontrivial topological properties that endow them with rich internal geometries — particularly through the Berry curvature, which encodes the geometric phase structure of Bloch electrons in momentum space. When incorporated into hydrodynamic descriptions, Berry curvature acts as an effective field that modifies the flow dynamics, leading to extended fluid models with new transport behavior and emergent geometries. These systems, with their nontrivial topological features, present compelling platforms for realizing dynamically evolving analog spacetime, where geometric and topological corrections naturally intertwine with fluid evolution. When embedded into hydrodynamic formulations, these corrections give rise to extended fluid models with enriched transport phenomena and novel emergent geometries. Such systems naturally admit nonlinear couplings between flow variables and geometric quantities, making them ideal candidates for hosting dynamically evolving analog spacetime.

In this work, we synthesize and advance prior developments in analog gravity and hydrodynamic models incorporating Berry curvature by formulating a nonlinear analog spacetime framework within a prototypical topological material. Starting from  Berry curvature modified fluid equations, we develop a fully nonlinear perturbation theory around a spherically symmetric, stationary background. By employing a quantity, which is a look-alike of mass accretion rate, as the perturbation variable, we derive a nonlinear wave equation that naturally defines a time dependent, effective acoustic metric.\\
Our framework retains all nonlinear contributions without truncation and enables a systematic perturbative expansion of metric corrections to arbitrary order. Notably, the dynamical metric encodes nonlinearities through its dependence on both background and fluctuating fields, and reproduces known linear results in the appropriate limit. Going beyond, it captures rich dynamics including oscillatory horizons, frequency-dependent causal structure, and evolving curvature.

Moreover, we numerically study these emergent spacetime in the context of graphene, a two-dimensional topological material whose electronic transport properties exhibit hydrodynamic behavior consistent with our fluid model. Tracking the analog horizon under high and low frequency perturbations, we uncover regimes of dynamical relaxation and horizon recession through ``low frequency" perturbations, with no direct counterpart in classical black hole physics. Importantly, our framework allows for the estimation of an effective Hawking temperature, which in realistic experimental regimes — e.g., graphene with typical carrier densities — lies in the range of {\bf{tens of microkelvins}}, well within reach of current detection techniques.\\
Overall, our results establish a topologically enriched, nonlinear analog gravity framework, offering insight into emergent spacetime dynamics and positioning topological materials as promising testbeds for probing phenomena such as horizon formation, back-reaction, and analog Hawking radiation.

Our paper is organized as follows. In Section \ref{sec:dynamics}, we introduce the spherically symmetric fluid system and describe our perturbation method, which leads to an acoustic spacetime of fully nonlinear perturbations of the mass accretion rate due to Berry curvature effects. Section \ref{sec:Hawking} explores the geometric configuration of the acoustic horizon and derives the associated Hawking temperature. Our numerical findings are presented in Section \ref{sec:numerical}. Section \ref{sec:background} briefly outlines the topological material system serving as the background fluid model in our study, while Sec. \ref{sec:zeroth} discusses the emergence of the acoustic horizon within the stationary background solution of our model. Subsequently, in Sec. \ref{sec:phase}, we conduct a phase space analysis of a test particle's trajectory propagating through this stationary background flow. In Sec. \ref{sec:Hawking-Numerics}, we numerically compute the Hawking temperature for our system, obtaining a value on the order of a ten of micro-kelvin. Furthermore, Sec. \ref{sec:perturbation} presents solutions up to second order for the inverse acoustic metrics, arising from time dependent perturbations with exponential damping in our fluid model. Additionally, we examine in detail on the fundamental assumptions for both high frequency (Sec. \ref{sec:highfreq}) and low frequency (Sec. \ref{sec:lowfreq}) perturbation regimes, along with their consequences for the acoustic horizon. Finally, we discuss a brief conclusion.

\section{Nonlinear perturbations of spherically symmetric radial flow}\label{sec:dynamics}

We study an extended fluid model with the Berry curvature effects \cite{Mitra:2021wjp}, whose semiclassical phase space formulation models electron dynamics in a magnetic Bloch band system, accounting for both the periodic crystal potential under an external magnetic field ${\bf{B}}$ and the non-trivial geometric effects arising from the Berry curvature ${\bf{\Omega}}({\bf {k}})$ in momentum space $\bf{k}$~\cite{Duval:2005vn,Gosselin:2006ht,Chang:1995zz}. The dynamics of such extended fluid models can be written as follows \cite{Mitra:2021wjp}.
\begin{equation}
    \dot{\rho}+\nabla \left( \frac{\rho \mathbf{v}}{\mathcal{A}}\right)=0\label{eq:conservation}~,
\end{equation}
\begin{equation}
    \dot{\mathbf{v}} + \frac{(\mathbf{v} \cdot \nabla)\mathbf{v}}{\mathcal{A}}+\frac{\nabla P}{\rho \mathcal{A}}=0\label{eq:Euler}~,
\end{equation}
where the fluid model is characterized by the density $\rho$ and pressure $P(\rho)$. For small Berry curvature $\bf{\Omega(k)}$, one can write the Berry curvature related term $\mathcal{A}$ with electronic charge $e$ as $\A({\bf{x,k}})=1+e(\bf{\O(k)}.\bf{B(x)})$ \cite{Mitra:2021wjp}.\\
\noindent
We note that one can solve Eqs.~\eqref{eq:conservation}, and \eqref{eq:Euler} in their steady-state forms, yielding velocity and density fields that depend solely on spatial coordinates, expressed as $v \equiv v_0(r)$ and $\rho \equiv \rho_0(r)$. In perturbation theory~\citep{pso80}, a common approach involves introducing small time-dependent radial perturbations to these stationary profiles, $v_0(r)$ and $\rho_0(r)$, followed by linearization of the perturbed terms. While effective for small-amplitude evolution, this linear treatment offers limited insight into the full temporal dynamics of the hydrodynamic flow. Consequently, the natural progression is to incorporate nonlinear effects into the perturbation framework. By systematically including higher order nonlinear terms, the perturbative analysis gradually approximates the complete time dependent behavior of global solutions, beginning from a specified stationary configuration at time $t=0$. In what follows, we adopt the approach of Ref.~\cite{Sen:2012rb}, where our analysis is based on the accretion rate variable with a constant Berry curvature term $\mathcal{A}$, defined in the spherically symmetric background, given by~\cite{pso80,TD}
\begin{eqnarray}
    F(r,t)=\frac{\rho v r^2}{\A}~,\label{eq:F}
\end{eqnarray}
where the non-vanishing radial component of the velocity vector of the fluid is given by $v$. The prescription for the time dependent radial perturbation of radial velocity and background density is given by
\begin{eqnarray}
    v(r,t)=v_0(r)+v^{\prime}(r,t)~,\no\\
    \rho(r,t)=\rho_0(r)+\rho^{\prime}(r,t)\label{eq:valocity-density}~.
\end{eqnarray}
Here the $\prime$ denotes a perturbation about a stationary background. In the stationary limit, both velocity and density fields acquire purely spatial profiles. The perturbation in $F$ is obtained while fully preserving the nonlinear aspects of the system, by implementing the perturbation method for variables $v$ and $\rho$ in Eq.~\eqref{eq:valocity-density} as
\begin{eqnarray}
    \frac{F^{\prime}}{F_0}=\frac{v^{\prime}}{v_0}+\frac{\rho^{\prime}}{\rho_0}+\frac{\rho^{\prime}v^{\prime}}{\rho_0 v_0}\label{eq:nonlinear-all}~.
\end{eqnarray}
This expression establishes a connection between the perturbed quantities $v^\prime$, $\rho^\prime$, and $F^\prime$. Here the constant $F_0$ can be interpreted as the Berry curvature induced matter flow rate, scaled by a factor of $4\pi\A$~\cite{Frank}, which can be expressed in terms of $v_0,~\rho_0$ and the constant term $\A$ as $F_0=(\rho_0 v_0 r^2)/\A$.\\

In the presence of constant Berry curvature term $\A$, the dynamics of the fluid in spherically symmetric background is given by the continuity equation as follows.
\begin{eqnarray}
    \frac{\p\r}{\p t}=-\frac{1}{r^2}\frac{\p}{\p r}\Big(\frac{\rho vr^2}{\A}\Big)\label{eq:continuity}~.
\end{eqnarray}
To proceed, we first relate $\rho^\prime$ and $F^\prime$ analytically. Applying the perturbation method on the above equation, we can have a direct relationship between $\rho^\prime$ and $F^\prime$ as the following.
\begin{eqnarray}
    \frac{\p\r^{\prime}}{\p t}=-\frac{1}{r^2}\frac{\p F^{\prime}}{\p r}~.\label{eq:rho-F}
\end{eqnarray}
Now taking the time derivative in Eq.~\eqref{eq:nonlinear-all} and using Eq.~\eqref{eq:rho-F}, we get a direct relation between $v^{\prime}$ and $F^{\prime}$ as
\begin{eqnarray}
    \frac{\p v^{\prime}}{\p t}=\frac{v}{F}\Big(\frac{\p F^{\prime}}{\p t}+\frac{v}{\A}\frac{\p F^{\prime}}{\p r}\Big)~.\label{eq:v-F}
\end{eqnarray} 
Here all orders of nonlinearity are retained in Eqs.~\eqref{eq:nonlinear-all}, \eqref{eq:rho-F}, and \eqref{eq:v-F}.\\

\noindent
To derive a fully nonlinear perturbation equation involving the constant Berry curvature term $\A$, we now turn to the Euler equation in spherically symmetric background. The inviscid Euler equation (Eq.~\eqref{eq:Euler}) can be written as 
\begin{eqnarray}
    \frac{\p v}{\p t}+\frac{v}{\A}\frac{\p v}{\p r}+\frac{1}{\A\r}\frac{\p P}{\p r}=0~.\label{eq:Euler-special}
\end{eqnarray}
The local pressure $P$ is expressed in terms of the density $\rho$, using a general polytropic relation, $P=K \rho^\gamma$, where $\gamma$, the polytropic exponent, lies within the range $1\leq\gamma\leq c_P/c_V$ and $K$, being a constant. This range is constrained by isothermal and adiabatic conditions, with $c_P$ and $c_V$ denoting the specific heat capacities of a fluid at constant pressure and volume, respectively. On the other hand, the local speed of sound $(c_s)$ is determined from the relation $c_s^2=\partial P / \partial \rho = \gamma K \rho^{\gamma - 1}$.\\

Now by utilizing the perturbation method outlined in Eq.~\eqref{eq:Euler-special} and computing its second order partial derivative with respect to time, we obtain the following key result:
\begin{eqnarray}
    \frac{\p^2 v^{\prime}}{\p t^2}+\frac{\p}{\p r}\Big(\frac{c_s^2}{\A\r}\frac{\p \r^{\prime}}{\p t}+\frac{v}{\A}\frac{\p v^{\prime}}{\p t}\Big)=0~.\label{eq:Euler-pert}
\end{eqnarray}
Here, all stationary terms are eliminated via time differentiation, differing from the conventional method of separating the stationary part of Eq.~\eqref{eq:Euler}.\\
Now from Eq.~\eqref{eq:Euler-pert}, by utilizing Eqs.~\eqref{eq:rho-F}, and \eqref{eq:v-F}, we can write a fully nonlinear perturbation equation, which can be expressed in a symmetric form, as the following.
\begin{eqnarray}
    \frac{\p}{\p t}\left[\frac{v}{F}\frac{\p F^{\prime}}{\p t}\right]+\frac{\p}{\p t}\left[\frac{v^2}{\A F}\frac{\p F^{\prime}}{\p r}\right]+\frac{\p}{\p r}\left[\frac{v^2}{\A F}\frac{\p F^{\prime}}{\p t}\right]\no\\
    +\frac{\partial}{\partial r} \left[ \frac{v^3}{\A^2 F} \frac{\partial F^{\prime}}{\partial r} - \frac{c_s^2 v}{\A^2F} \frac{\partial F^{\prime}}{\partial r} \right] = 0~.\label{eq:pert}
\end{eqnarray}
This symmetric structure allows for a compact representation as
\begin{eqnarray}
    \partial_\mu \big[g^{\mu\nu} \partial_\nu F^{\prime}\big] = 0~;~~~~~~~~(\mu,\nu=0,1)~.\label{eq:compact}
\end{eqnarray}
This nonlinear wave equation--arising in the presence of Berry curvature via the term $\A$--will serve as the basis for our subsequent perturbative analysis. The nonlinearity is entirely incorporated in the metric components, $g_{\mu\nu}$, which depend on the exact field variables $v,~c_s,~F$ and $\A$, rather than solely on their stationary background counterparts \cite{Visser:1997ux,Schutzhold:2002rf}. This departs from conventional linearized analog spacetime descriptions and highlights the genuinely dynamical nature of the effective geometry.\\
Finally, from Eq.~\eqref{eq:pert}, we can define the components of the inverse acoustic metric as the following.
\begin{equation}\label{eq:matrix}
    g^{\mu \nu} = \frac{\A}{\rho r^2}
\begin{pmatrix}
    1 & \frac{F}{\rho r^2} \\
    \frac{F}{\rho r^2} & \Big(\frac{F^2}{\rho^2 r^4} - \frac{c_s^2}{\A^2}\Big) 
\end{pmatrix}~. 
\end{equation}
Before proceeding forward, it is important to emphasize that the analog spacetime be described by a metric $G_{\mu \nu}$, where the fluctuations obey the equation of a massless scalar field in this spacetime as  
\begin{equation}  
\frac{1}{\sqrt{-G}} \partial_{\mu} \left( \sqrt{-G} \, G^{\mu \nu} \partial_{\nu} F' \right) = 0~.\label{eq:wave}  
\end{equation}  
The conversion of the metric components from Eq.~\eqref{eq:compact} to Eq.~\eqref{eq:wave} is well-defined in spacetime with more than two dimensions. Given that our system exhibits spherical symmetry, the transformation from Eq.~\eqref{eq:compact} to Eq.~\eqref{eq:wave} arises from the relation $G^{\mu \nu} = \sqrt{-g}g^{\mu \nu}$, where $g$ denotes the determinant of the four dimensional metric. Irrespective of how the additional dimensions are incorporated, the components of $G^{\mu \nu}$ remain proportional to $g^{\mu \nu}$ and vice-versa, scaled by an overall factor dependent on the metric determinant. Since this scaling leaves both the causal structure and the dynamical properties unchanged, we choose to work with the unique two dimensional inverse metric $g^{\mu \nu}$ in the presence of a Berry curvature term $\A$, whose components are given by Eq.~\eqref{eq:compact}.\\
Therefore, the effective acoustic metric takes the following form.
\begin{eqnarray}
    ds^2&=&g_{\mu\nu}dx^{\mu}dx^{\nu}\nonumber\\
    &=&\frac{\rho r^2}{\A}\Big(1-\frac{\A^2F^2}{\rho^2 r^4 c_s^2}\Big)dt^2+\frac{2\A F}{c_s^2}dtdr-\frac{\A \rho r^2}{c_s^2}dr^2~.\nonumber\\
    \label{eq:metric}
\end{eqnarray}
We now examine the perturbative expansions of $F$ and $\rho$ around a known stationary solution up to the $n^{\text{th}}$-order. Suppose $\zeta(r,t)$ represents a field encompassing all dependent variables of the fluid system--such as density, the accretion rate incorporating the Berry curvature term, components of the inverse acoustic metric, fluid velocity, sound speed, and other relevant quantities. The $n^{\text{th}}$-order expansion of such a field $\zeta(r,t)$ can be expressed as
\begin{equation}  
\zeta(r,t) = \zeta_0(r) + \sum_{l=1}^{n}\epsilon^k \zeta_k(r,t) \equiv \zeta_0(r) + \zeta'(r,t)~,\label{eq:perturbative}  
\end{equation}  
where $\zeta_0$ corresponds to the stationary solution, and the dimensionless perturbation parameter $\epsilon$ determines the strength of the deviation. This expansion remains valid in a perturbative regime provided that $\epsilon \, \vert \zeta_{l+1} \vert / \vert \zeta_{l} \vert < 1$ for all $l = 0, \dots, (n-1)$. It is noted that Eq.~\eqref{eq:perturbative} is similar with Eq.~\eqref{eq:valocity-density} for the velocity and density of the fluid, respectively.\\

Now to derive the $n^{\text{th}}$-order solutions for $F$ and $\rho$, we begin by substituting their perturbative expansions, as defined in Eq.~\eqref{eq:perturbative}, into Eqs.~\eqref{eq:conservation} and \eqref{eq:compact}. The next step involves gathering terms with corresponding powers of $\epsilon$ and solving for the respective coefficients. The zeroth order coefficient ($\epsilon^0$) automatically satisfies the stationary solution of the system.\\
From the first order coefficient ($\epsilon$) obtained via Eq.~\eqref{eq:continuity}, we derive the following relation:
\begin{eqnarray}
   \frac{\partial \rho_1}{\partial t} = -\frac{1}{r^2}\frac{\partial F_1}{\partial r}~, \label{eq:zeroth_rho}
\end{eqnarray}   
while Eq.~\eqref{eq:compact} yields\\
\begin{eqnarray}
   \partial_\mu \left( g^{\mu\nu}_{(0)} \partial_\nu F_1(r,t) \right) = 0~. \label{eq:zeroth_metric}
\end{eqnarray}
Consequently, the inverse metric components of the zeroth order effective acoustic spacetime can be expressed as follows.
\begin{eqnarray}
    &g^{tt}_{(0)} = \frac{\mathcal{A}}{r^2 \rho_0}~, \quad g^{tr}_{(0)} \equiv g^{rt}_{(0)} = \frac{F_0 \mathcal{A}}{r^4 \rho_0^2}~,\nonumber \\
    &g^{rr}_{(0)} = \frac{F_0^2 \mathcal{A}}{\rho_0^3 r^6} - \frac{c^2_{s0}}{\mathcal{A} \rho_0 r^2}~.\label{eq:g_up0}
\end{eqnarray}

The inverse metric $g^{\mu\nu}_{(0)}$ is entirely determined by the stationary background solution, and its components coincide with the known stationary analog spacetime structure incorporating the Berry curvature term. On this background, linear fluctuations in the mass accretion rate propagate accordingly.\\
The corresponding acoustic metric $g_{\mu\nu}^{(0)}(r)$ can be expressed using Eq.~\eqref{eq:g_up0} as
\begin{eqnarray}
    &g_{tt}^{(0)}=\frac{\rho r^2}{\A}\Big(1-\frac{\A^2F_0^2}{\rho^2c_s^2r^4}\Big)~,~~g_{tr}^{(0)}\equiv g_{rt}^{(0)}=\frac{\A F_0}{c_s^2}~,\nonumber\\
    &g_{rr}^{(0)}=\frac{\A\rho r^2}{c_s^2}.\label{eq:g_down0}
\end{eqnarray}  

% ===================== Second-order perturbations =====================

Proceeding to second order perturbations from coefficients of $\epsilon^2$, we now derive the corresponding evolution equations as follows.
\begin{eqnarray}
   \frac{\p\r_2}{\p t}=-\frac{1}{r^2}\frac{\p F_2}{\p r}~,\label{eq:first_rho}\\
   \p_\mu\Big(g^{\mu\nu}_{(0)}\p_\nu F_2(r,t)\Big)&=-\p_\mu\Big(g^{\mu\nu}_{(1)}\p_\nu F_1(r,t)\Big)~\label{eq:first_metric}.
\end{eqnarray}
The first order correction $g_{(1)}^{\mu \nu}$ (denoted as $g^{\mu\nu}_1$ in their respective graphical representations) to the inverse metric contains contributions from both $F_1$ and $\rho_1$:
\begin{align}
    &g^{tt}_{(1)} = \frac{\A}{r^2 \rho_0}\left(-\frac{\rho_1}{\rho_0}\right)~,~~g^{tr}_{(1)}\equiv g^{rt}_{(1)} = \frac{\A F_0}{r^4 \rho_0^2}\left(\frac{F_1}{F_0} - 2 \frac{\rho_1}{\rho_0}\right), \notag\\
    &g_{(1)}^{rr} =\frac{\A F_0^2}{r^6 \rho_0^3} \left(2 \frac{F_1}{F_0} - 3 \frac{\rho_1}{\rho_0}\right)-\frac{c_{s0}^2}{\A\rho_0 r^2}\left(\frac{\rho_1}{c_{s0}^2}\frac{\partial c_{s}^2}{\partial \rho}\Bigg\vert_{\rho_0} - \frac{\rho_1}{\rho_0}\right)~,\label{eq:g1}
\end{align}
where $\frac{\partial c_{s}^2}{\partial \rho}\Big\vert_{\rho_0}$ represents the differentiation of $c_s^2$ with respect to $\rho$, which is evaluated at $\rho_0$.

% ===================== Third-order perturbations =====================

The third order fluctuation equation from coefficients of $\epsilon^3$ includes source terms from both first and second order inverse metric corrections. Therefore for the third order fluctuation equations, we can write
\begin{eqnarray}
   \frac{\p\r_3}{\p t}=-\frac{1}{r^2}\frac{\p F_3}{\p r}~,\label{eq:second_rho}\\
   \p_\mu\Big(g^{\mu\nu}_{(0)}\p_\nu F_3(r,t)\Big)&=-\p_\mu\Big(g^{\mu\nu}_{(2)}\p_\nu F_1(r,t)\Big)\nonumber\\
   &-\p_\mu\Big(g^{\mu\nu}_{(1)}\p_\nu F_2(r,t)\Big)~\label{eq:second_metric}.
\end{eqnarray}
The components of $g^{\mu \nu}_{(2)}$ (denoted as $g^{\mu\nu}_2$ in their respective graphical representations) contain quadratic and mixed products of first order fluctuations as well as explicit second order contributions:
\begin{align}
    &g^{tt}_{(2)} = \frac{\A}{r^2 \rho_0}\left(\left(\frac{\rho_1}{\rho_0}\right)^2 -\frac{\rho_2}{\rho_0}\right)~,\no\\
    &g_{(2)}^{tr}\equiv g^{rt}_{(2)} = \frac{\A F_0}{r^4\rho_0^2}\left(\frac{F_2}{F_0} - 2 \left(\frac{\rho_2}{\rho_0} + \frac{F_1}{F_0}\frac{\rho_1}{\rho_0}\right) + 3 \left(\frac{\rho_1}{\rho_0}\right)^2\right)~,\no\\
    &g^{rr}_{(2)} = \frac{\A F_0^2}{r^6 \rho_0^3} \Bigg[\left(\frac{F_1}{F_0}\right)^2 + 2 \frac{F_2}{F_0} - 3 \frac{\rho_2}{\rho_0} + 6\left(\left(\frac{\rho_1}{\rho_0}\right)^2 - \frac{F_1}{F_0}\frac{\rho_1}{\rho_0}\right)\Bigg] \no\\
    &- \frac{c_{s0}^2}{\A\rho_0 r^2}\Bigg[\frac{\rho_2}{c_{s0}^2}\frac{\partial c_{s}^2}{\partial \rho}\Bigg\vert_{\rho_0} + \frac{\rho^2_1}{2 c_{s0}^2}\frac{\partial^2 c_{s}^2}{\partial \rho^2}\Bigg\vert_{\rho_0} - \frac{\rho_1^2}{c_{s0}^2\r_0} \frac{\partial c_{s}^2}{\partial \rho}\Bigg\vert_{\rho_0}\no\\
    &- \frac{\rho_2}{\rho_0} + \left(\frac{\rho_1}{\rho_0}\right)^2\Bigg]~.
    \label{eq:g2}
\end{align}

It is important to observe that Eq.~\eqref{eq:compact} enables us to understand Eqs.~\eqref{eq:zeroth_metric}, \eqref{eq:first_metric}, and \eqref{eq:second_metric} as describing a combined third order perturbation of the Berry curvature induced mass accretion rate $F_0 + \epsilon F_1 + \epsilon^2 F_2$ that propagates through a second order effective acoustic background characterized by the inverse metric $g_{(0)}^{\mu \nu} + \epsilon g_{(1)}^{\mu \nu}+\epsilon^2 g_{(2)}^{\mu\nu}$, without any loss of generality. This perspective offers an accurate description of the perturbation equations up to second order, while simultaneously permitting us to interpret their solutions as a unified massless disturbance moving through an effective acoustic spacetime.

Additionally, let us mention that the emergence of the effective acoustic metric, as captured by the nonlinear perturbation equation, provides us with the framework for analyzing the dynamical response of the system under time dependent perturbations. With explicit expressions for the inverse metric components up to several orders of perturbation, we can investigate how these behave in realistic settings. In the later sections, we will focus on the numerical simulations that show how the background flow and its perturbative changes evolve when Berry curvature effects come into play. Before moving on to the numerical section, one important step remains, which is to determine the location of the horizon in this analog setup and derive the corresponding expression for the Hawking temperature of the system.

%{With explicit expressions for the inverse metric components obtained across multiple perturbative orders, we are now equipped to explore their behavior in realistic scenarios. In particular, we turn to numerical simulations that reveal how the background flow, as well as its perturbative corrections, evolve in the presence of Berry curvature effects. These results, presented in the next section, will illuminate the dynamical structure of the analog spacetime and its sensitivity to both high and low frequency fluctuations.}

\section{Horizon location and Hawking Temperature of the analog System}\label{sec:Hawking}
In order to calculate the Hawking temperature for our acoustic metric, we start with the background velocity flow whose normal component is essential in this case, which is written as
\begin{equation}
    \vec{v}_{\perp}(r) = \frac{b}{r}\hat{r}~\label{eq:velo_normal},
\end{equation}
where its magnitude is $|\vec{v}_{\perp}(r)|=b/r$ \cite{Barcelo:2011fc}. Here $b$ is constant, dimensionally equivalent to the specific angular momentum parameter. In our study, we are dealing with $b<0$, which means a future acoustic horizon (acoustic black hole).

Then, from Eq.~\eqref{eq:F}, the Berry curvature induced matter flux is then given by
\begin{equation}
    F(r) =\frac{b \rho(r) r}{\A}~.\label{eq:matter_flux}
\end{equation}
On the other hand, the expression for $g^{rr}$ given in Eq.~\eqref{eq:matrix} reveals that $g^{rr}$ becomes zero when  
\begin{equation}  
    r_H^2 = \frac{F\A}{\rho c_s}~, \label{eq:horizon1}  
\end{equation}  
where $r_H$ represents the radius of the acoustic horizon. Furthermore, as evident from Eq.~\eqref{eq:F}, the condition $v(r_H) = c_s$ holds at the location $r = r_H$, indicating that the horizon is situated precisely at the critical point where the fluid's flow velocity equals its local sound speed. Consequently, $g^{rr}$ takes negative values in the subsonic regime $(v < c_s)$ and positive values in the supersonic regime $(v > c_s)$.\\
Therefore we can also write down the location of the acoustic horizon in terms of the local sound speed $c_s$, and the specific angular momentum parameter $b$ as follows.
\begin{equation}
    r_H = \frac{|b|}{c_s}~.\label{eq:horizon2}
\end{equation}

Now, to compute the Hawking temperature, we begin with the definition of the surface gravity for acoustic metrics \cite{Barcelo:2011fc}
\begin{equation}
	\kappa = \frac{1}{2} \frac{\partial}{\partial r} \left[ \frac{\rho(r)\, r^2}{\A} \left(c_s^2 - v(r)^2 \right) \right] \bigg|_{r = r_H}~.\label{eq:kappa1}
\end{equation}
Substituting the velocity profile $v(r) = b/r$ and evaluating at the horizon, where $v(r_H)^2= c_s^2$, we find that only the third term in the expansion contributes. Therefore the expression for the surface gravity for acoustic metrics can be written as
\begin{equation}
    \kappa = \frac{\rho(r_H) \, b^2}{\A r_H}~.\label{eq:kappa2}
\end{equation}

Thus, the Hawking temperature is given by
\begin{equation}
    T_H = \frac{\hbar\kappa}{2\pi k_{B}} = \frac{\hbar\rho(r_H) \, b^2}{2\pi k_{B} \A r_H}~.\label{eq:T_H}
\end{equation}
This result confirms that the analog Hawking temperature in our extended fluid model depends on the fluid density at the horizon $\rho(r_H)$, the specific angular momentum parameter $b$, and the Berry curvature term $\A$. The numerical value of the Hawking temperature will be computed for some particular parameter values of this system and will be discussed in detail in a subsequent section.

\section{Numerical analysis}\label{sec:numerical}

Now in this section, we begin by analyzing how the background density of our extended fluid model varies with the spatial coordinate $x$. To illustrate this, we present a characteristic profile of the background density as a function of radial distance for a specific topological insulator graphene system. Subsequently, we examine the lowest order solutions of the acoustic metric and its corresponding inverse metric components. Finally, we perform a numerical investigation of the accretion rate variable $F$ when a Berry curvature term $\A$ is present. This includes studying the density $\rho$ of our extended fluid model and all inverse acoustic metric components, focusing on the scenario where the accreting flow solution exhibits an exponentially decaying temporal perturbation in topological insulator systems. Our analysis treats the hydrodynamics of charge carriers within an extended fluid framework that incorporates Berry curvature effects.

\subsection{Background Density Distribution}\label{sec:background}

We obtain the background density distribution from first principles quantum mechanical calculation on a graphene unit cell. Our density-functional theory (DFT) calculations for structural relaxation were carried out in a plane-wave basis with projector-augmented wave (PAW) potentials~\cite{blochl} as implemented in the Vienna Ab-initio Simulation Package (VASP)~\cite{kresse, kresse01}.
In all our DFT relaxation calculations, we chose as exchange-correlation functional the Generalised Gradient Approximation implemented following the Perdew-Burke-Ernzerhof (PBE) prescription~\cite{pbe}.
For ionic relaxations using the VASP package, the internal positions of the atoms were allowed to relax until the forces became less than 0.005\,eV/$A^0$. An energy cutoff of 600\,eV and an 8$\times$8$\times$4 Monkhorst–Pack $k$-points mesh provided good convergence of the total energy.

\begin{figure}[H]
    \begin{center}
    \includegraphics[width=1.0\linewidth]{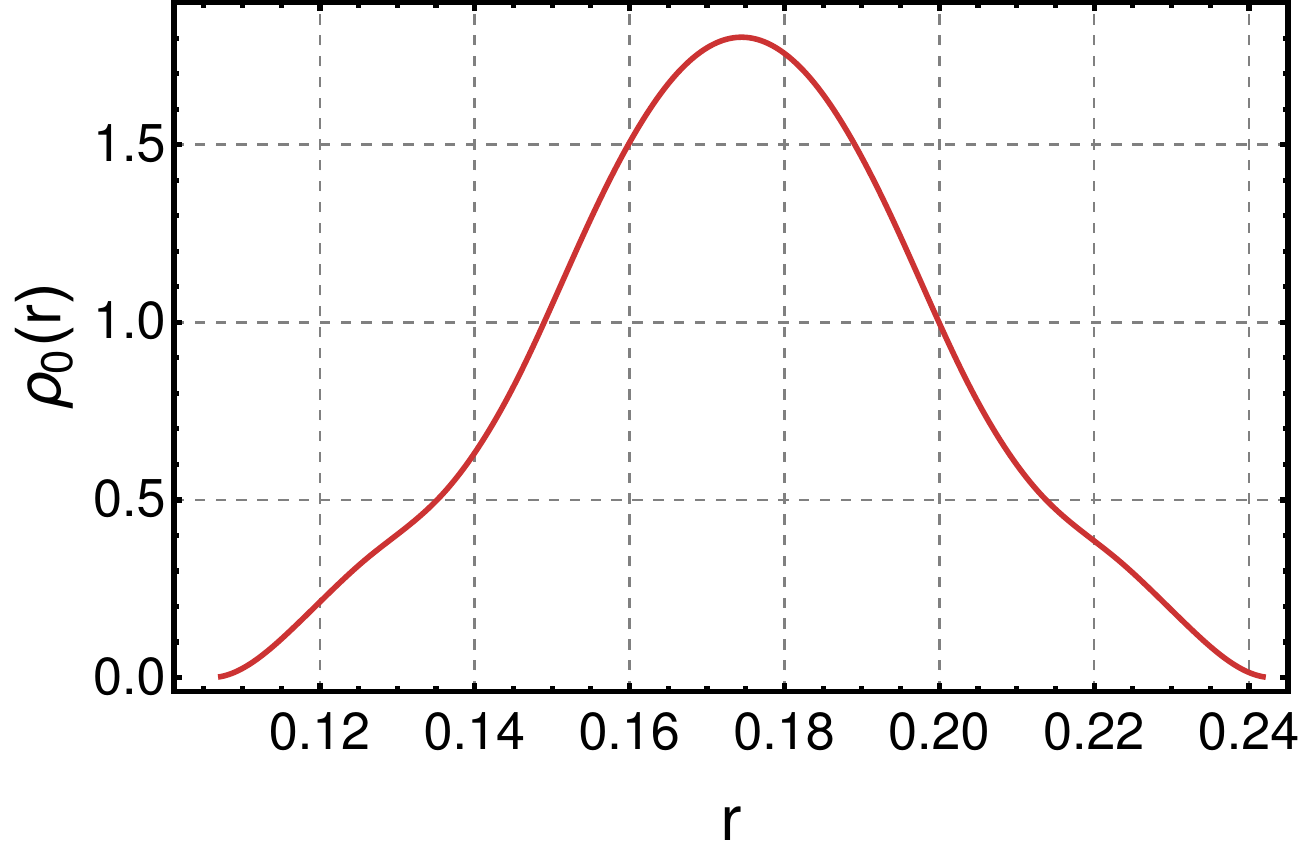}
    \end{center}
    \caption{Plot illustrating the background charge density distribution $\rho_0$ of graphene for our extended fluid model. In this case, the radial coordinate is defined as $r \equiv \sqrt{2}x$, considering the equivalence between the $x$ and $y$ directions in the two dimensional graphene plane, and we plot on the radial coordinate range $0.1068478~\rm{m}\leq r \leq 0.2421708~\rm{m}$, as depicted in the figure.}\label{f:rho_final}
\end{figure}

\subsection{Zeroth Order Acoustic Metric and Horizon Structure}\label{sec:zeroth}

Next, for accretion flows within our extended fluid framework, the equation of state can generally be chosen as either isothermal or isentropic. In this work, we adopt the isothermal assumption for the flow, which leads to the relation $P = K \rho$, where $K$ is a measure of the fluid’s constant specific entropy. Under this isothermal approximation, the local sound speed remains uniform throughout the flow. As previously discussed in Sec.~\ref{sec:Hawking}, we consider the fluid velocity profile following the form presented in Eq.~\eqref{eq:velo_normal}.\\

\begin{widetext}
        \begin{figure*}
	\begin{center} 
        \includegraphics[width=1.0\linewidth]{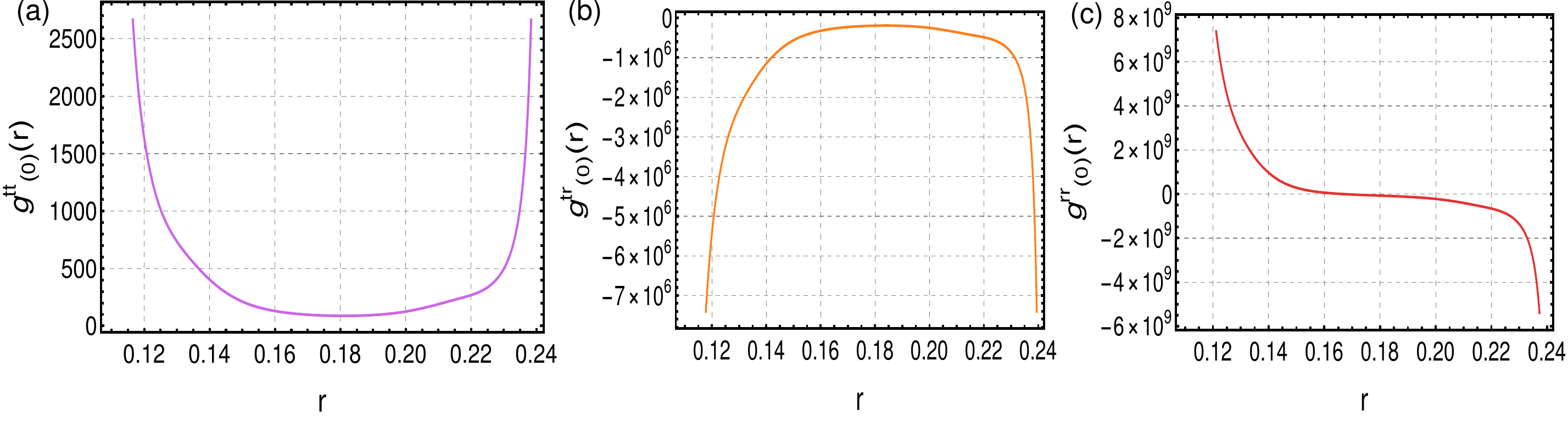}\label{f:gtt0_up}
        \begin{minipage}{\textwidth}
        \caption{Plots displaying all components of the inverse acoustic metric (Eq.~\eqref{eq:g_up0}), derived from our extended fluid model for a black hole with the parameter $b = -1999.37~\rm{m^2/s}$ and a constant sound speed $c_s = 12000.0~\rm{m/s}$. Fig.~$2(a)$ illustrates the stationary behavior of the inverse acoustic metric for the $tt$ component. Fig.~$2(b)$ depicts the stationary variation of the inverse acoustic metric in the $tr \equiv rt$ components. Fig.~$2(c)$ shows the stationary variation of the inverse acoustic metric for the $rr$ component. The horizon position for the stationary flow, where $g^{rr}_{(0)}(r) = 0$, is located at $r = 0.166614~\rm{m}$.}\label{f:g_up0_all}
        \hrulefill
        \end{minipage}
        \end{center}
        \begin{center}
        \includegraphics[width=1.0\linewidth]{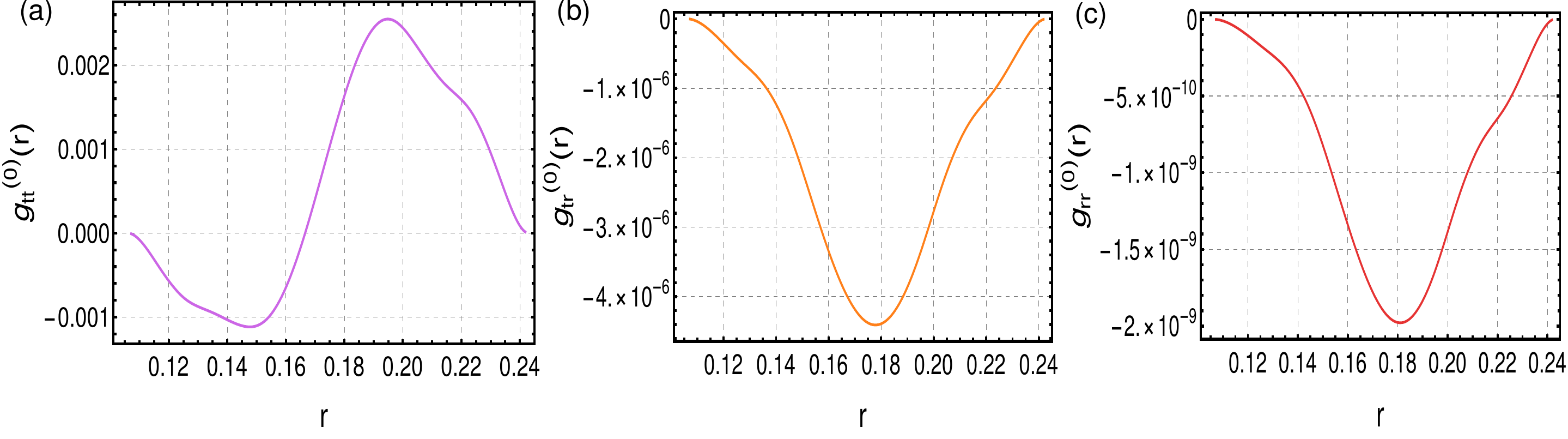}\label{f:grr0_down}
        \begin{minipage}{\textwidth}
        \caption{Plots displaying all components of the acoustic metric (Eq.~\eqref{eq:g_down0}), derived from our extended fluid model for a black hole with the parameter $b = -1999.37~\rm{m^2/s}$ and a constant sound speed $c_s = 12000.0~\rm{m/s}$. Fig.~$3(a)$ illustrates the stationary behavior of the acoustic metric for the $tt$ component. Fig.~$3(b)$ depicts the stationary variation of the acoustic metric in the $tr \equiv rt$ components. Fig.~$3(c)$ shows the stationary variation of the acoustic metric for the $rr$ component. The horizon position for the stationary flow, where $g_{tt}^{(0)}(r) = 0$, is located at $r = 0.166614~\rm{m}$.}\label{f:g_down0_all}
        \end{minipage}
        \end{center}
        \end{figure*}
    \end{widetext}

The transonic solution further determines the computation of the lowest order acoustic metric components and their inverses through Eqs.~\eqref{eq:g_down0} and \eqref{eq:g_up0}, respectively. Notably, these zeroth order effective acoustic metric components serve as the background for first order accretion rate perturbations incorporated with the Berry curvature term. At this zeroth order level, all components of both the acoustic metric and its inverse exhibit time independence, with their solutions graphically represented in Figs.~\ref{f:g_up0_all} and \ref{f:g_down0_all}.\\
Analysis of the zeroth order inverse acoustic metric components from Eq.~\eqref{eq:g_up0} reveals that the inverse acoustic horizon, where $g^{rr}_{(0)} = 0$, occurs at $r_H=0.166614~\rm{m}$. Figure~$2(c)$ demonstrates that $g^{rr}_{(0)}$ assumes positive values for $r<r_H$ and negative values for $r>r_H$. In contrast, the remaining inverse metric components show consistent behavior: $g^{tt}_{(0)}$ remains strictly positive while $g^{tr}_{(0)}$ maintains a negative definite character throughout the flow regime (as evidenced by Figs.~$2(a)$ and $2(b)$, respectively).\\
Conversely, the acoustic metric component $g^{(0)}_{tt}$ displays a negative within the horizon ($r<r_H$) and positive outside ($r>r_H$), as shown in Fig.~$3(a)$. This behavior precisely mirrors the opposite pattern of $g^{rr}_{(0)}$, consistent with theoretical expectations. The other metric components $g^{(0)}_{tr}$ and $g^{(0)}_{rr}$ maintain negative definite values across the entire flow regime, as illustrated in Figs.~$3(b)$ and $3(c)$ respectively.

The effective acoustic spacetime at zeroth order emerges from perturbations in the accreting flow solution within our extended fluid model, which includes a Berry curvature term $\A$. We will elaborate on this in detail in the subsequent subsections. However, before moving forward, it is worthwhile to examine how a probe particle's dynamics evolve in the stationary background derived from our extended fluid model. In what follows, in the next subsection, we will explore the possible characteristics of the phase space dynamics for a probe particle located near the acoustic horizon.

\subsection{Phase-Space Dynamics in a Stationary Background}\label{sec:phase}

To further explore the characteristics of our analog black hole geometry, we examine the trajectory of a massive test particle within the stationary background obtained from our extended fluid model, which includes the influence of Berry curvature. This approach enables us to analyze particle dynamics near the acoustic horizon from a dynamical perspective. We consider the same background flow configuration, where the radial fluid velocity is described by Eq.~\eqref{eq:velo_normal}. Meanwhile, the background density $\rho_0(r)$ is derived from an analytical fit to the charge distribution data of a topological insulator graphene (see Fig.~\ref{f:rho_final}), providing a realistic model of the effective medium.

To study the particle's motion, we employ the Hamiltonian framework. The time independent (stationary) nature of the background flow in Eq.~\eqref{eq:g_down0} ensures energy conservation. Consequently, we express the standard covariant dispersion relation for a massive probe particle of mass $m$ as follows:
\begin{equation}
    g^{\mu\nu} p_\mu p_\nu = -m^2~,\label{eq:dispersion}
\end{equation}
where $p_{\mu}$ represents the four-momentum vector, which in our case corresponds to the canonical momenta $(p_t,~p_r)$. Owing to the static symmetry of the acoustic metric geometry, a timelike Killing vector exists, allowing us to define energy as $E = -p_t$. Using the zeroth order effective inverse acoustic metric $g^{\mu\nu}_{(0)}$ from Eq.~\eqref{eq:g_up0}, we construct the Hamiltonian via the covariant dispersion relation given in Eq.~\eqref{eq:dispersion}. 

    \newpage

The resulting Hamiltonian, expressed in terms of the radial momentum $p_r$, radial coordinate $r$, and mass $m$, takes the following form:
\begin{eqnarray}
    E=\frac{bp_r\pm\sqrt{4c_s^2r^2p_r^2-3b^2p_r^2-4\A m^2r^4\r_0^2}}{2\A r}~.\label{eq:energy}
\end{eqnarray}
Notably, the conserved Hamiltonian in the above equation is independent of the stationary background density $\r_0(r)$ for a massless probe particle.

The equations of motion are then derived using Hamilton's equations:
\begin{eqnarray}
    &\dot{r} &= \frac{\partial E}{\partial p_r}\nonumber\\
    &=&\frac{b}{2\A r}+\frac{(4c_s^2r^2p_r-3b^2p_r)}{2\A r\sqrt{4c_s^2r^2p_r^2-3b^2p_r^2-4\A m^2r^4\r_0^2}}~,\nonumber\\
    \label{eq:coupled_r}\\
    &\dot{p}_r &= -\frac{\partial E}{\partial r}\nonumber\\
    &=&\frac{bp_r}{2\A r^2}+\frac{\sqrt{4c_s^2r^2p_r^2-3b^2p_r^2-4\A m^2r^4\r_0^2}}{4\A^2 r^3}\nonumber\\
    &-&\frac{(4c_s^2p_r^2 r-8\A m^2\r_0^2r^3-4\A m^2r^4\r_0\r_0')}{2\A r\sqrt{4c_s^2r^2p_r^2-3b^2p_r^2-4\A m^2r^4\r_0^2}}~,\label{eq:coupled_pr}
\end{eqnarray}
where $\r_0'(r)$ denotes the derivative of the stationary background density $\r_0(r)$ with respect to $r$. To obtain the phase-space trajectory, these coupled differential equations are numerically solved for a unit probe massive particle with an initial value of $r$, but the canonical momenta $p_r$ is not free and can be found from Eq.~\eqref{eq:energy} for a fixed unit energy.

\begin{figure}[H]
    \begin{center}
    \includegraphics[width=1.0\linewidth]{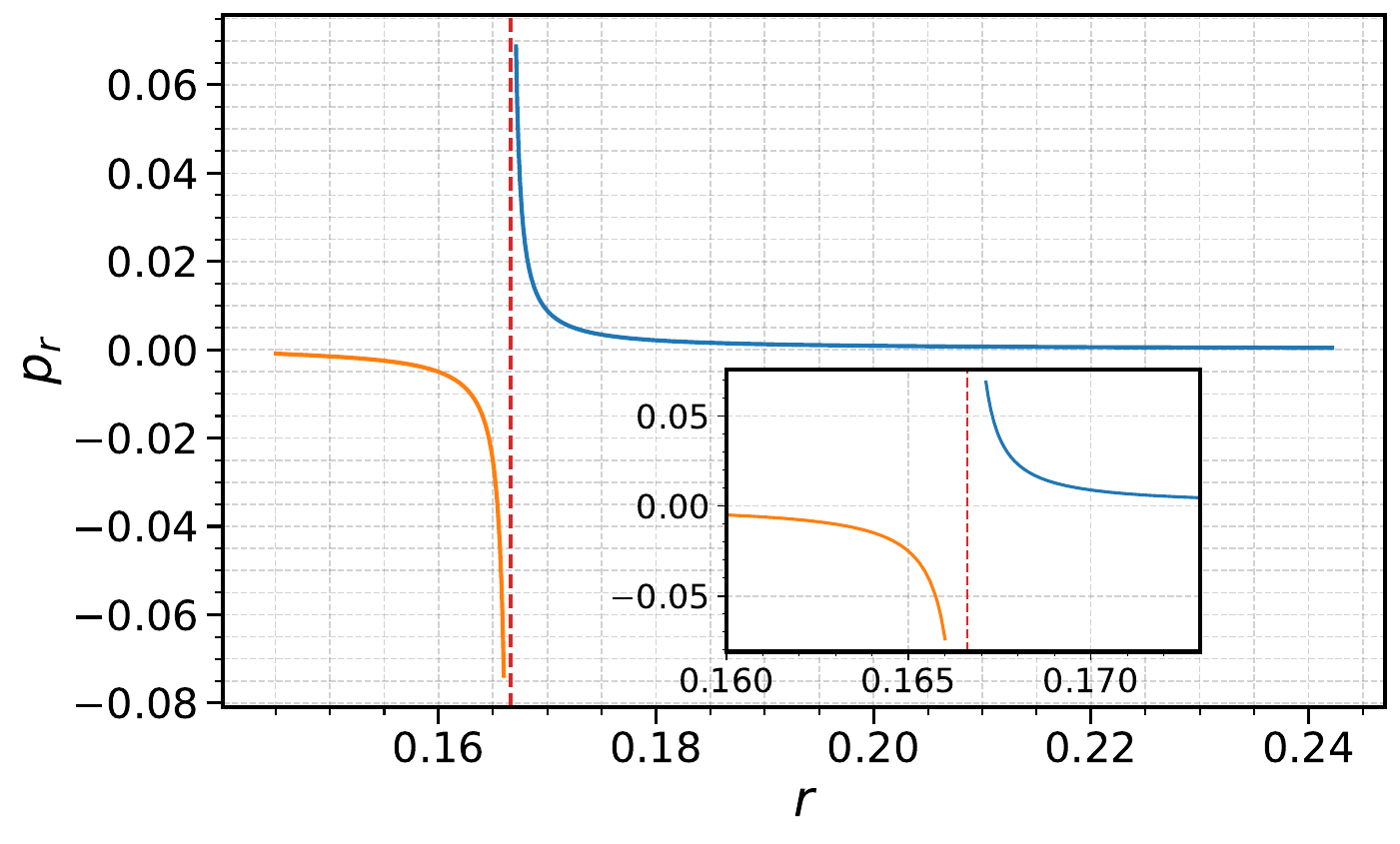}
    \end{center}
    \caption{Plot illustrating the phase-space trajectory of a unit massive probe particle, whose motion is considered in the stationary background solution of graphene for our extended fluid model. From this figure, it is clearly seen that the particle moves near to $ r= 0.166614~\rm{m}$, its radial conjugate momentum increases exponentially, reaching its high peak as $r$ attains $0.166614~\rm{m}$. On the other hand, beyond $r=0.166614~\rm{m}$, behavior of probe particle's motion is just opposite. This occurrence enables us to find out the position of the horizon, which precisely corresponds to the theoretical expectation at $|b|/c_s=0.166614~\rm{m}\equiv r_H$. Here the red dotted line shows the location of the acoustic horizon $r\equiv r_H=0.166614~\rm{m}$.}\label{f:ps}
\end{figure}

Fig.~\ref{f:ps} displays our result on the behavior of the radial momentum $p_r$ to the radial coordinate $r$ of a massive probe particle. A striking feature appears near $r = 0.166614~\rm{m}$, which matches the theoretically predicted location of the acoustic horizon computed from the condition $g^{rr}_{(0)}(r_H) = 0$ in the above section. As particle approaches this radius, its radial momentum increases rapidly and reaches a sharp peak very close to this location $r= 0.166614~\rm{m}$. This behavior indicates a dynamical instability or sudden change in the trajectory—consistent with the physical picture of a horizon acting as a one-way membrane. Interestingly, similar behavior has been observed in other studies \cite{Dalui:2024xer} of near-horizon physics, where particles show a ``sudden change" or ``instability" in their motion as they get close to the horizon. This characteristic is one of the near-horizon features which has been studied recently in some black hole contexts \cite{Dalui:2019esx,Dalui:2020qpt,Dalui:2021tvy}. Beyond the horizon (i.e., for $r < 0.166614~\rm{m}$), the particle’s momentum quickly decreases, suggesting it is being drawn inward into the supersonic region of the flow, much like an in-falling object in the vicinity of a gravitational black hole.

The phase-space trajectory provides a clear dynamical signature of the acoustic horizon. The near-horizon behavior of such a probe particle is a rectangular hyperbola (see the inset plot of Fig.~\ref{f:ps}). Moreover the sudden rise and fall in the particle's momentum near the critical radius closely mirrors the trapping behavior found in black hole physics. This analysis not only strengthens the analogy with real black holes but also confirms horizon formation dynamically in our nonlinear, Berry curvature modified fluid system.

\subsection{Numerical Estimate of the Hawking Temperature}\label{sec:Hawking-Numerics}

To estimate the Hawking temperature of our analog system, we use the expression of Eq.~\eqref{eq:T_H}. Now, for the realizable example we take the horizon location from Fig.~\ref{f:g_up0_all}, $r_H\approx 0.166614~\rm{m}$, which corresponds to the theoretical value of $r_H = |b|/c_s$ for the specific angular momentum parameter $b = -1999.37~\rm{m^2/s}$ and the constant sound speed $c_s = 12000.0~\rm{m/s}$. Now, from Fig.~\ref{f:rho_final}, the background density at this location is approximately $\rho(r_H)\equiv 1.711766~\rm{C/m^2}$. Substituting these values into the temperature formula and setting the constant Berry curvature term $\A = 5$, we obtain
\begin{eqnarray}
    T_H &\approx& \frac{1.711766 \times (-1999.37)^2}{2\pi \times 5\times 1.666140\times 10^{-1}}\times 7.638226\times 10^{-12}~ \rm K~,\nonumber\\
    &\approx& 9.98532 \times 10^{-6} ~\rm K~,
\end{eqnarray}
where $\hbar/k_{B}=7.638226\times10^{-12}~\rm{K~s}$. The value of $T_{H}$ here represents the analog Hawking temperature generated by the effective horizon in our extended fluid model of graphene and it lies in the micro-kelvin regime. In this context, it is important to highlight that several studies, as referenced in \cite{Blencowe:2020ygo,Hrishit1,Hrishit2}, have similarly reported numerical estimates of the Hawking temperature within the micro-kelvin scale.

\subsection{Perturbations on Accretion Rate Variable, Density and the Components of Inverse Acoustic Metric}\label{sec:perturbation}

In this section, we will examine the temporal evolution of radial perturbations affecting the Berry curvature-induced accretion rate variable $F$, charge density $\rho$, and all components of the inverse acoustic metric $g^{\alpha\beta}$ within our extended fluid model framework. Through numerical analysis, we study the behavior of $F$, $\rho$, and the inverse acoustic metric components under time dependent perturbations that decay exponentially as $e^{-i\omega t}$, applied to our accreting flow solution in a topological insulator. Our investigation focuses on the flow dynamics while accounting for fluctuations up to second order, providing comprehensive insights into the perturbed system.\\
The perturbation is applied starting at $t = 0~\rm{s}$ and analyzed until $t = 30~\rm{s}$, a duration significantly longer than the spatial scale, as the `large time' limit. This selection guarantees that damping effects remain meaningful without dominating the background solution. Additionally, the inner and outer spatial boundaries of the accreting fluid for our system enable us to examine two distinct frequency regimes. Perturbations with `high frequencies' correspond to wavelengths comparable to or smaller than the inner boundary so that in our extended fluid model framework, $\omega_{\text{high}} \ge 7.0566\times 10^{5}~\rm{rad/s}$. On the other hand, `low frequency' perturbations are associated with wavelengths exceeding the radius of the outer boundary of the flow. Therefore for our system we have $\omega_{\text{low}} \le 3.11343\times 10^{5}~\rm{rad/s}$. In our whole analysis, we adopt $\omega_{\text{high}} = 10^{6}~\rm{rad/s}$ and $\omega_{\text{low}} = 10^{4}~\rm{rad/s}$.

\subsubsection{High frequency perturbations}\label{sec:highfreq}

First, for the Berry curvature incorporated accretion rate perturbations, in solving Eq.~\eqref{eq:zeroth_metric}, we assume the trial solution of $F_1(r,t)$ of the form:
\begin{eqnarray}
    F_1(r,t)=g_{\o}(r)e^{-i\o t}~.\label{eq:F1_form}
\end{eqnarray}
Therefore using Eq.~\eqref{eq:F1_form}, from Eq.~\eqref{eq:zeroth_metric} one can have the following differential equation of $g_{\o}(r)$.
\begin{eqnarray}
    &g^{rr}_{(0)}~g''_{\o}+\big(g^{rr}_{(0)}\big)^{\prime}~g'_{\o}-\o^2 g^{tt}_{(0)}~g_{\o}\nonumber\\
    &-i\o\Big(2g^{tr}_{(0)}~g'_{\o}+\big(g^{tr}_{(0)}\big)^{\prime}~g_{\o}\Big)=0~,\label{eq:F1diff}
\end{eqnarray}
where the $\prime$ denotes derivative with respect to radial coordinate $r$. To solve the given differential equation, one can express $g_{\omega}(r)$ as a superposition of real and imaginary solutions.

\begin{widetext}
        \begin{figure*}
	\begin{center} 
        \includegraphics[width=1.0\linewidth]{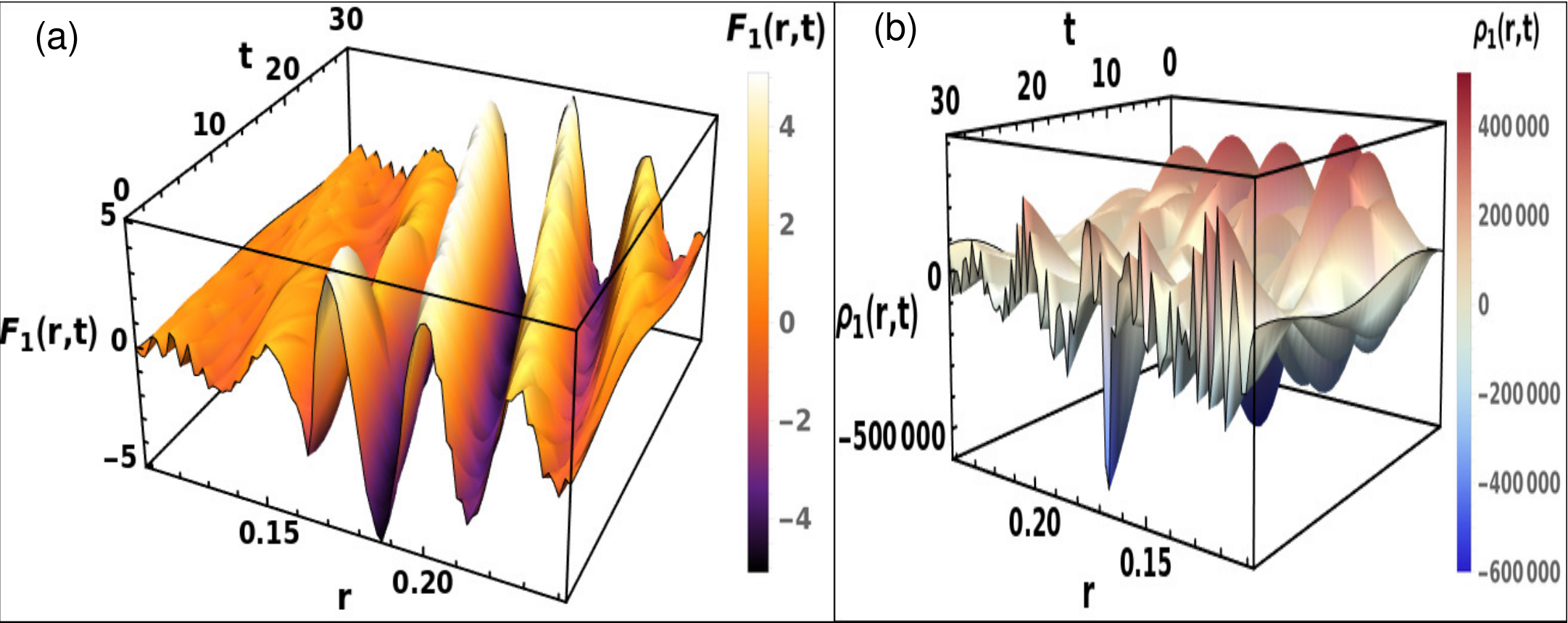}
        \begin{minipage}{\textwidth}
        \caption{The $3D$ graphical representations depict the solutions for the first order perturbed accretion rate and density within our extended fluid model, for high frequency perturbations taking $\omega_{\text{high}} = 10^6~\rm{rad/s}$. Fig.~$5(a)$ demonstrates the first order perturbed behavior of the accretion rate variable $F_1(r,t)$, which includes the influence of the Berry curvature term over space and time. Meanwhile, Fig.~$5(b)$ shows the corresponding first order perturbation in the charge density profile, $\rho_1(r,t)$ over space and time.}\label{f:F1rho1_h}
        \hrulefill
        \end{minipage}
        \end{center}
        \begin{center}
        \includegraphics[width=1.0\linewidth]{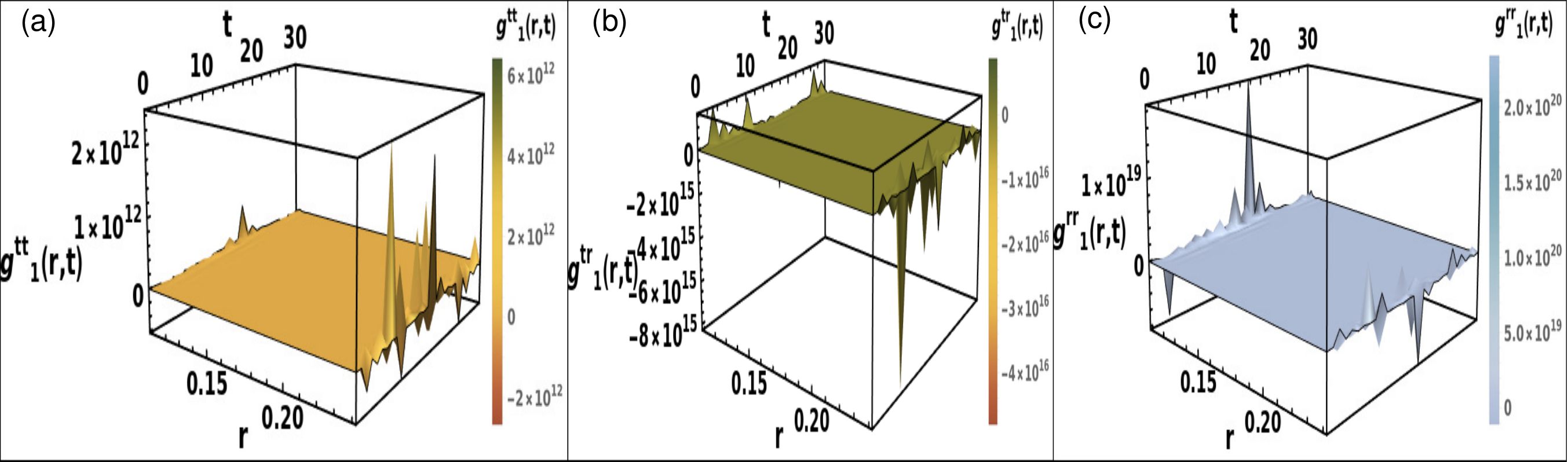}
        \begin{minipage}{\textwidth}
        \caption{Plots displaying all components of the first order inverse acoustic metric (Eq.~\eqref{eq:g1}), derived from our extended fluid model with high frequency perturbation $\omega_{high}=10^6~\rm{rad/s}$. Fig.~$7(a)$ illustrates the spacetime behavior of the first order perturbed inverse acoustic metric for the $tt$ component. Fig.~$7(b)$ depicts the spacetime variation of the first-order perturbed inverse acoustic metric in the $tr \equiv rt$ components. Fig.$7(c)$ shows the spacetime variation of the first order perturbed inverse acoustic metric for the $rr$ component.}\label{f:g1all_h}
        \end{minipage}
        \end{center}
        \end{figure*}
    \end{widetext}

The complete real solution for $F_1(r,t)$ is illustrated in Fig.~\ref{f:F1rho1_h}(a). In the subsonic regime (where $r > 0.166614~\rm{m}$), we observe the interaction between growing and decaying modes, accompanied by an amplitude that increases near the horizon from subsonic to supersonic flow in all time. This phenomenon has been previously documented in traveling wave solutions and is a characteristic feature of subsonic mass accretion rate perturbations \cite{Petterson:1980}. Notably, within our extended fluid framework, this behavior (as depicted in Fig.~\ref{f:F1rho1_h}(a)) is similar with recent studies on first order perturbations in the mass accretion rate for Bondi accretion flows \cite{Fernandes:2021gkf, Fernandes:2022bwo}. However, once the horizon is crossed, the interaction amplitude between the growing and decaying modes begins to dimish in the supersonic region.

To obtain the real solution for the charge density $\rho_1(r,t)$ within our extended fluid model system, we now utilize the complete real solution of $F_1(r,t)$. Consequently, using the continuity equation given in Eq.~\eqref{eq:zeroth_rho}, the numerically computed solution for $\rho_1(r,t)$ is illustrated in Fig.~\ref{f:F1rho1_h}(b). The charge density fluctuation $\rho_1(r,t)$ is significant both in the subsonic and supersonic region, which can be expected due to the constructive interference of $F_1(r,t)$ in both of this domain (as shown in Fig.~\ref{f:F1rho1_h}(a)). Additionally, it should be emphasized that $\rho_1(r,t)$ reaches its peak negative fluctuation in the vicinity of the acoustic horizon. Given that the perturbations possess wavelengths comparable to or shorter than the inner boundary of the accretor, our fluid model demonstrates that these disturbances can penetrate into the accretor, sustaining density variations. Furthermore, the amplitude of both growing and decaying modes exhibits stronger fluctuations during the initial time as opposed to the `late-time' phase.

Next to investigate the behavior of the first order perturbed inverse acoustic metric components with a time dependent profile, we incorporate the stationary solutions for the Berry curvature induced mass accretion rate $F_0(r)$ and the background charge density $\rho_0(r)$ from our extended fluid model, along with the first order perturbation solutions of $F_1(r,t)$ and $\rho_1(r,t)$, into the inverse acoustic metric component expressions provided in Eq.~\eqref{eq:g1}. The resulting first order perturbed inverse acoustic metric components are illustrated in Fig.~\ref{f:g1all_h}.

Fig.~\ref{f:g1all_h} clearly illustrates that the first order fluctuations in all inverse acoustic metric components are significant only at the inner and outer boundaries of the flow regime. Specifically, the fluctuations in $g^{tt}_{(1)}(r,t)$ attain their maximum positive amplitude at the outer boundary ($r=0.2421708~\rm{m}$) of the subsonic flow during the initial time stage around $t \approx 8~\rm{s}$ (as shown in Fig.~\ref{f:g1all_h}(a)). A comparable pattern, though with negative amplitude, is seen for the fluctuations in $g^{tr}_{(1)}(r,t)$ (displayed in Fig.~\ref{f:g1all_h}(b)). Notably, the fluctuation in $g^{rr}_{(1)}(r,t)$ achieves its peak positive magnitude at the inner boundary ($r=0.1068478~\rm{m}$) of the supersonic flow near $t \approx 18~\rm{s}$, which contrasts with the behavior of $g^{tt}_{(1)}(r,t)$ and $g^{tr}_{(1)}(r,t)$ observed in the subsonic region. Importantly, apart from these two boundaries, the spacetime fluctuations of all first order perturbed inverse acoustic metrics effectively become negligible within the framework of our topological insulator system.

We now address the solution of the second order perturbation equations governing the Berry curvature-induced mass accretion rate variable and charge density, as presented in Eq.~\eqref{eq:first_metric} and Eq.~\eqref{eq:first_rho}, respectively. To determine $F_2(r,t)$, we utilize the numerical solutions obtained from the first order mass accretion rate $F_1(r,t)$, the first order inverse acoustic metric components $g^{\mu\nu}_{(1)}$, and the stationary background inverse acoustic metric components $g^{\mu\nu}_{(0)}$.

\begin{widetext}
        \begin{figure*}
	\begin{center} 
        \includegraphics[width=1.0\linewidth]{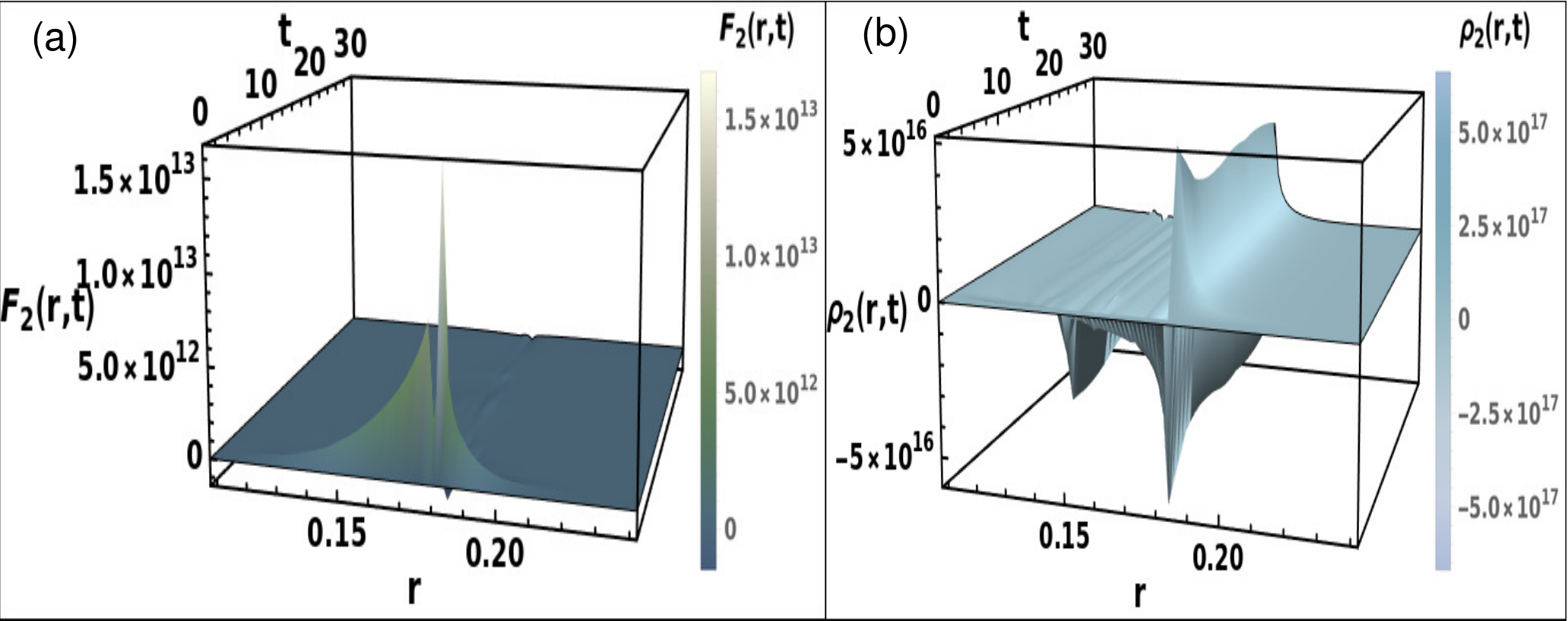}
        \begin{minipage}{\textwidth}
        \caption{The $3D$ graphical representations depict the solutions for the second order perturbed accretion rate and density within our extended fluid model, for high frequency perturbations taking $\omega_{\text{high}} = 10^6~\rm{rad/s}$. Fig.~$6(a)$ demonstrates the second order perturbed behavior of the accretion rate variable $F_2(r,t)$, which includes the influence of the Berry curvature term over space and time. Meanwhile, Fig.~$6(b)$ shows the corresponding second order perturbation in the charge density profile, $\rho_2(r,t)$ over space and time.}\label{f:F2rho2_h}
        \hrulefill
        \end{minipage}
        \end{center}
        \begin{center}
        \includegraphics[width=1.0\linewidth]{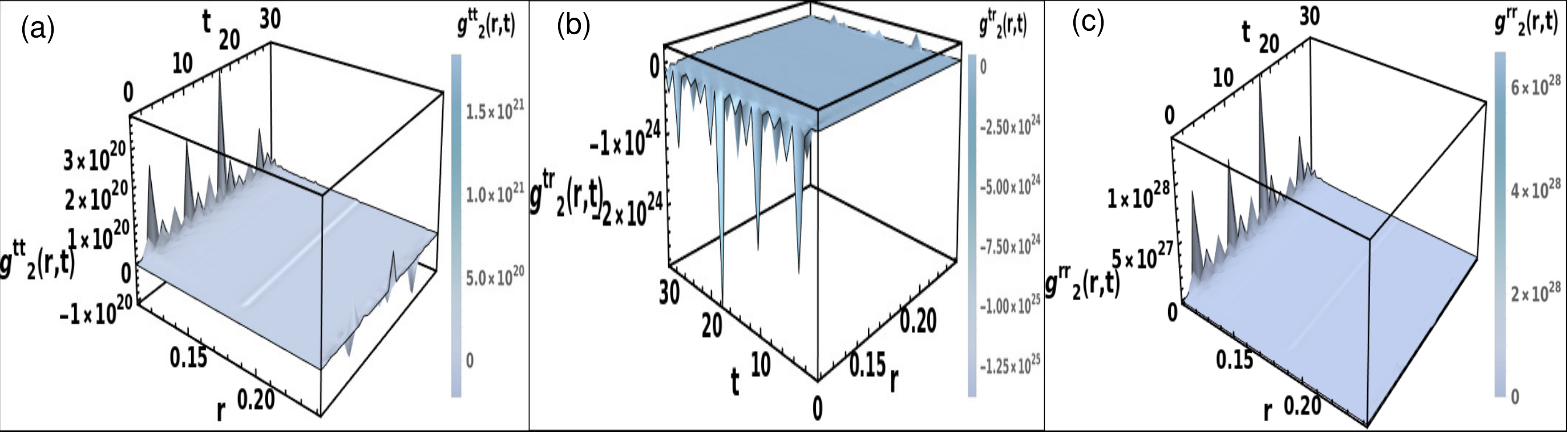}
        \begin{minipage}{\textwidth}
        \caption{Plots displaying all components of the second order inverse acoustic metric (Eq.~\eqref{eq:g2}), derived from our extended fluid model with high frequency perturbation $\omega_{high}=10^6~\rm{rad/s}$. Fig.~$8(a)$ illustrates the spacetime behavior of the second order perturbed inverse acoustic metric for the $tt$ component. Fig.~$8(b)$ depicts the spacetime variation of the second order perturbed inverse acoustic metric in the $tr \equiv rt$ components. Fig.~$8(c)$ shows the spacetime variation of the second order perturbed inverse acoustic metric for the $rr$ component.}\label{f:g2all_h}
        \end{minipage}
        \end{center}
        \end{figure*}
    \end{widetext}

This approach guarantees alignment with the first order perturbation solution while preserving the initial condition of exponential temporal decay. Similarly, for the second order density perturbation $\rho_2(r,t)$, we rely exclusively on the solutions of the second order perturbed mass accretion rate $F_2(r,t)$ to perform numerical integration. The resulting solutions for $F_2(r,t)$ and $\rho_2(r,t)$ are illustrated in Figs.~\ref{f:F2rho2_h}(a) and \ref{f:F2rho2_h}(b), respectively.

By analyzing the second order perturbations in the Berry curvature induced mass accretion rate and charge density alongside their first order counterparts (as illustrated in Fig.~\ref{f:F1rho1_h}), we observe that the second order perturbations exhibit a more amplified versions of the first order solutions in terms of their magnitudes. Specifically, the fluctuations in mass accretion are significantly enhanced in the vicinity of the acoustic horizon ($0.160162~\rm{m}\leq r \leq 0.199893~\rm{m}$) beginning at initial time $t = 0~\rm{m}$. These fluctuations attain their peak positive amplitude close to $r\approx 0.1829139~\rm{m}$ in the subsonic region at initial time stage $t=0~\rm{s}$. Outside the near-horizon region, the perturbations in both subsonic and supersonic flow regimes remain relatively flat across all time.

    \newpage
    
On the other hand, the density perturbation grows even more significantly at this order, as evident from the comparison between Fig.~\ref{f:F2rho2_h}(b) and Fig.~\ref{f:F1rho1_h}(b). Given that the mass accretion rate fluctuation $F_2(r,t)$ exhibits its largest magnitude in the subsonic region, it can be deduced that the corresponding density solution $\rho_2(r,t)$ includes minor corrections that start to emerge in the subsonic region at this order. Similar to the behavior observed for $F_2(r,t)$, the second order density fluctuation becomes more prominent in the vicinity of the horizon $(0.160162~\rm{m} \leq r \leq 0.199893~\rm{m})$. Furthermore, within the subsonic region at approximately $r \approx 0.1829139~\rm{m}$, the second order density fluctuation—considering both positive and negative domains—reaches its highest peak at the initial time $t = 0~\rm{s}$. A similar behavior, but with negative amplitude, is observed in the supersonic region around $r \approx 0.152861~\rm{m}$, though its magnitude is comparatively smaller than the former case.

    \newpage

The inverse metric components at second order are obtained by solving Eq.~\ref{eq:g2}, yielding the results shown in Fig.~\ref{f:g2all_h}. When comparing these plots with the first order solutions in Fig.~\ref{f:g1all_h}, it becomes evident that the observed trends become more pronounced at this order, with an overall relevant fluctuations emerging at the inner accretor boundary for the supersonic flow region.

For the component $g^{tr}_{(2)}$ (shown in Fig.~\ref{f:g2all_h}(b)), the fluctuation grows more negative in the supersonic region near the inner accretor boundary at $r\approx 0.1068478~\rm{m}$, reaching its maximum peak magnitude around $t\approx 18~\rm{s}$, unlike its first order counterpart $g^{tr}_{(1)}$. The second order perturbed solutions $g^{tt}_{(2)}$ exhibit stronger fluctuations at the inner accretor boundary for supersonic flow, opposite to the fluctuation behavior observed in the first order solutions $g^{tt}_{(1)}$. Notably, $g^{tt}_{(2)}$ achieves its highest positive value near $t\approx 18~\rm{s}$. Furthermore, in the case of $g^{rr}_{(2)}$, fluctuations are significant only at the inner accretor boundary for supersonic flow, with the largest positive magnitude occurring around $t\approx 18~\rm{s}$. Additionally, we find that the second order perturbed solutions for $g^{rr}_{(2)}$ generate exclusively positive fluctuations at the inner boundary throughout the entire time evolution.

Based on our comprehensive analysis of both first and second order perturbed solutions for the Berry curvature induced accretion rate variable, charge density, and all the inverse acoustic metric components under high frequency perturbations, we find that the maximum values show an order-of-magnitude increase when moving from first to second order solutions. Consequently, the suppression effect when transitioning from $\epsilon$ to $\epsilon^2$ proves particularly significant in our investigation.

\subsubsection{Low frequency perturbations}\label{sec:lowfreq}

\begin{widetext}
        \begin{figure*}
	\begin{center} 
        \includegraphics[width=1.0\linewidth]{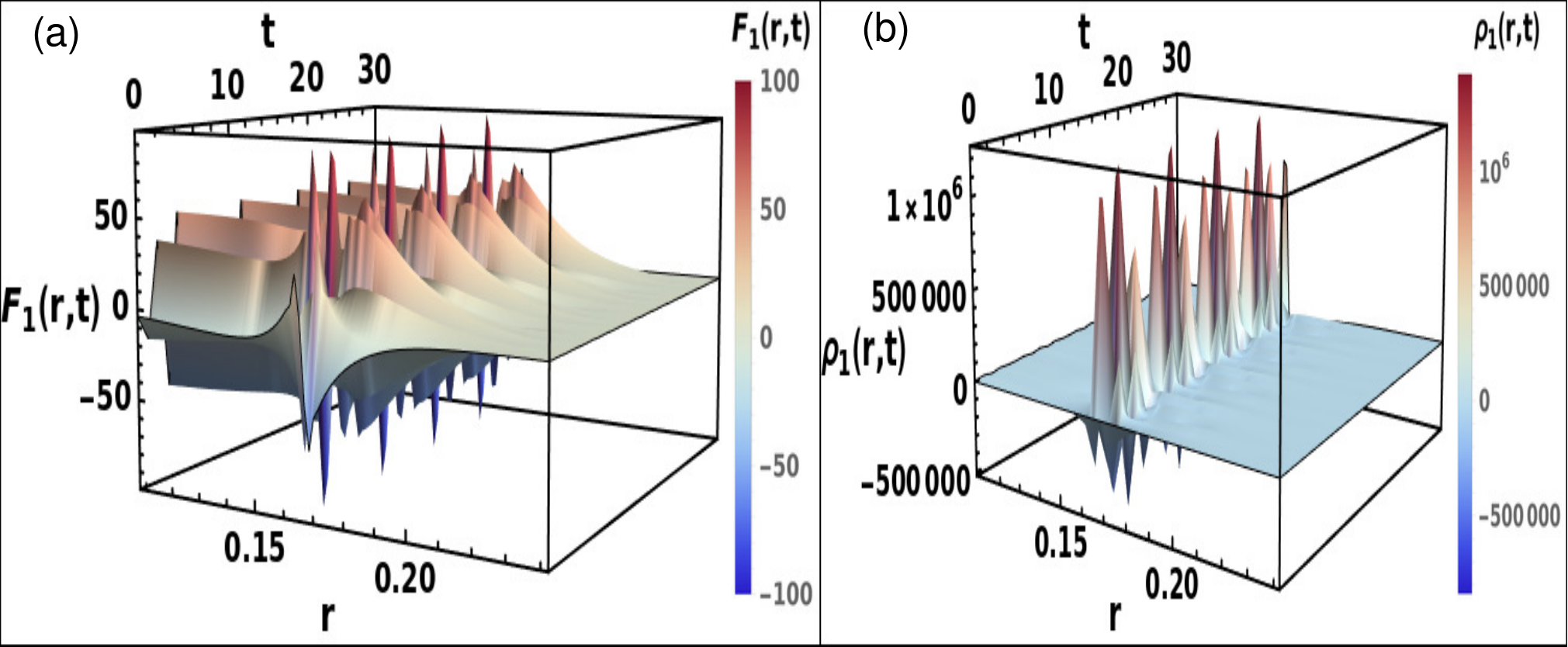}
        \begin{minipage}{\textwidth}
        \caption{The $3D$ graphical representations depict the solutions for the first order perturbed accretion rate and density within our extended fluid model, for low frequency perturbations with $\omega_{\text{low}} = 10^{4}~\rm{rad/s}$. Fig.~$9(a)$ demonstrates the first order perturbed behaviour of the accretion rate variable $F_1(r,t)$, which includes the influence of the Berry curvature term over space and time. Meanwhile, Fig.~$9(b)$ shows the corresponding first order perturbation in the charge density profile, $\rho_1(r,t)$ over space and time.}\label{f:F1rho1_l}
        \hrulefill
        \end{minipage}
        \end{center}
        \begin{center}
        \includegraphics[width=1.0\linewidth]{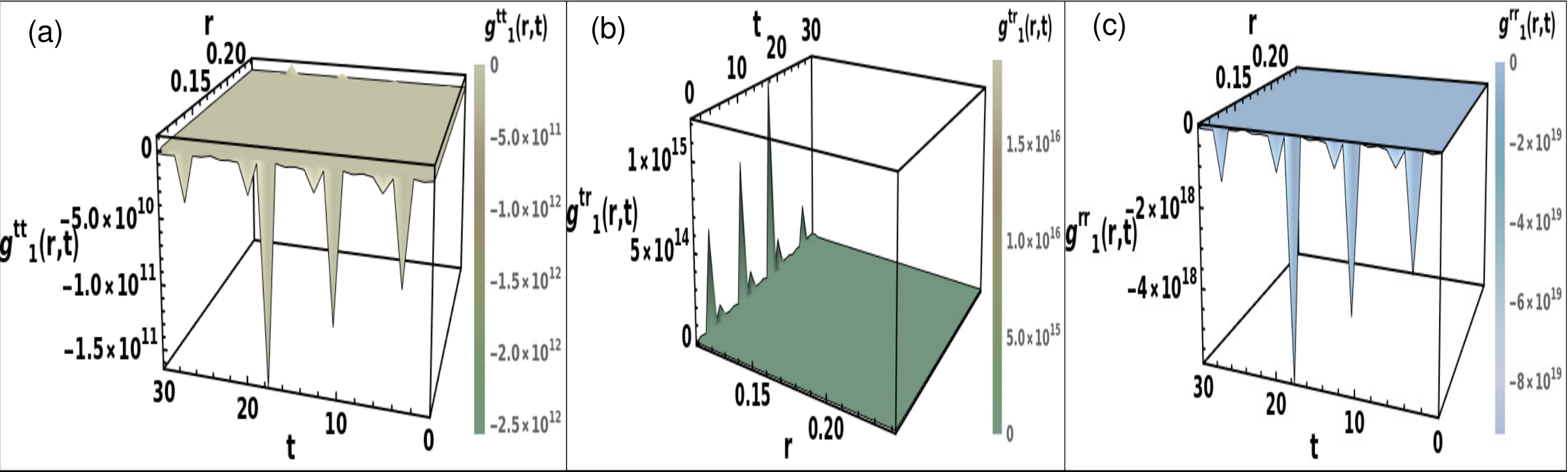}
        \begin{minipage}{\textwidth}
        \caption{Plots displaying all components of the first order inverse acoustic metric (Eq.~\eqref{eq:g1}), derived from our extended fluid model with low frequency perturbation $\omega_{low}=10^{4}~\rm{rad/s}$. Fig.~$11(a)$ illustrates the spacetime behaviour of the first order perturbed inverse acoustic metric for the $tt$ component. Fig.~$11(b)$ depicts the spacetime variation of the first order perturbed inverse acoustic metric in the $tr \equiv rt$ components. Fig.$11(c)$ shows the spacetime variation of the first-order perturbed inverse acoustic metric for the $rr$ component.}\label{f:g1all_l}
        \end{minipage}
        \end{center}
        \end{figure*}
    \end{widetext}

We now examine low frequency perturbations that decay exponentially over time. For this purpose, we numerically solve Eq.~\ref{eq:F1diff} with $\omega_{\textrm{low}} = 10^{4}~\rm{rad/s}$ over the time interval $t = 0~\rm{s}$ to $t = 30~\rm{s}$. The full real spacetime solution for the Berry curvature induced accretion rate, $F_1(r,t)$, is presented in Fig.~\ref{f:F1rho1_l}(a). The solution displays consistent oscillations between subsonic and supersonic flow throughout the entire time. Significantly, the first order perturbed mass accretion rate shows its largest spike-like variation close to the acoustic horizon radius. This finding underscores the crucial role of the acoustic horizon in shaping the dynamics of $F_1(r,t)$ when subjected to low frequency perturbations in our generalized fluid model system.

A similar spike-like behavior is also observed in the solution of the first order perturbed density $\rho_1(r,t)$ for low frequency perturbations. As shown in Fig.~\ref{f:F1rho1_l}(b), the solution becomes notably significant only near the acoustic horizon position at $r_H \approx 0.166614~\rm{m}$. In this region, more spike-like fluctuations exhibit over the entire time duration near the acoustic horizon radius, particularly when compared to the low frequency perturbed $F_1(r,t)$ solution. This characteristic further validates the existence of the horizon within our fluid model system under such low-frequency perturbations.

Next, to analyse the characteristics of the first order perturbed inverse acoustic metric components with a time dependent profile in the low frequency regime, we use the stationary solutions for the Berry curvature induced mass accretion rate $F_0(r)$ and the background charge density $\rho_0(r)$ from our extended fluid model. Along with these, we include the first order low frequency perturbation solutions $F_1(r,t)$ and $\rho_1(r,t)$ into the inverse acoustic metric component expressions given in Eq.~\eqref{eq:g1}. The resulting first-order perturbed inverse acoustic metric components are depicted in Fig.~\ref{f:g1all_l}.

Fig.~\ref{f:g1all_l} clearly demonstrates that for our extended fluid model system, the first-order low-frequency perturbations in all inverse acoustic metric components are dominantly noticeable within the supersonic flow region near the inner accretor boundary radius. In contrast to the high frequency perturbation solutions of $g^{tt}_{(1)}(r,t)$ and $g^{rr}_{(1)}(r,t)$ at this order (depicted in Figs.~\ref{f:g1all_h}(a) and \ref{f:g1all_h}(c), respectively), the perturbations here exhibit negative corrections. Specifically, for the components $g^{tt}_{(1)}(r,t)$ and $g^{rr}_{(1)}(r,t)$ (illustrated in Figs.~\ref{f:g1all_l}(a) and \ref{f:g1all_l}(c), respectively), the most pronounced negative amplitude occurs around time $t \approx 18~\rm{s}$. Conversely, unlike the high frequency perturbation of $g^{tr}_{(1)}(r,t)$, its low frequency counterpart reaches its peak positive value at approximately the same time, $t \approx 18~\rm{s}$.

Based on these observations, we can infer that the persistent negative corrections near the inner accretor boundary at $r \approx 0.1068478~\rm{m}$ suggest the presence of a receding acoustic horizon in our studied topological material system under the consideration of low frequency perturbation. Furthermore, it is worth noting that a similar phenomenon of a receding acoustic horizon has recently been explored through low frequency perturbations in the context of Bondi flow solutions by Fernandes \textit{et al.} \cite{Fernandes:2021gkf,Fernandes:2022bwo}.

\begin{widetext}
        \begin{figure*}
	\begin{center} 
        \includegraphics[width=1.0\linewidth]{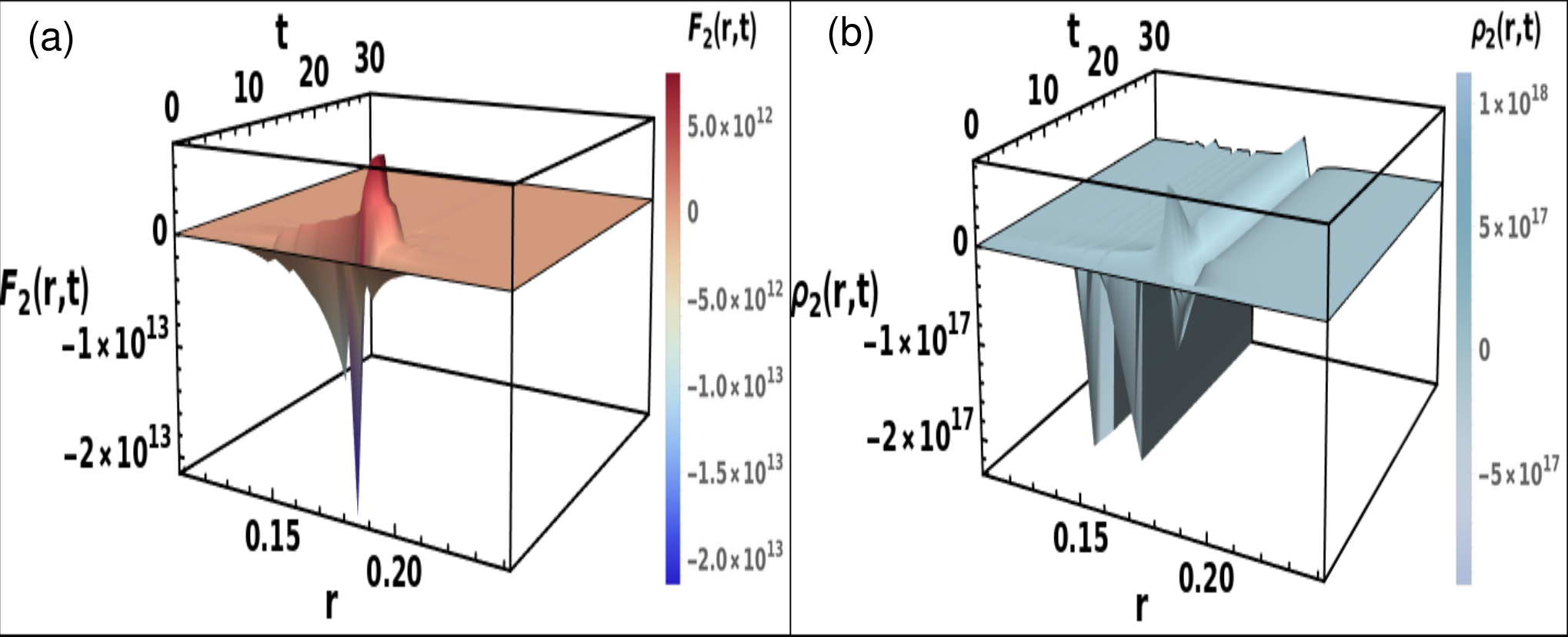}
        \begin{minipage}{\textwidth}
        \caption{The $3D$ graphical representations depict the solutions for the second order perturbed accretion rate and density within our extended fluid model, for low frequency perturbations with $\omega_{\text{low}} = 10^{4}~\rm{rad/s}$. Fig.~$10(a)$ demonstrates the second order perturbed behavior of the accretion rate variable $F_2(r,t)$, which includes the influence of the Berry curvature term over space and time. Meanwhile, Fig.~$10(b)$ shows the corresponding second-order perturbation in the charge density profile, $\rho_2(r,t)$ over space and time.}\label{f:F2rho2_l}
        \hrulefill
        \end{minipage}
        \end{center}
        \begin{center}
        \includegraphics[width=1.0\linewidth]{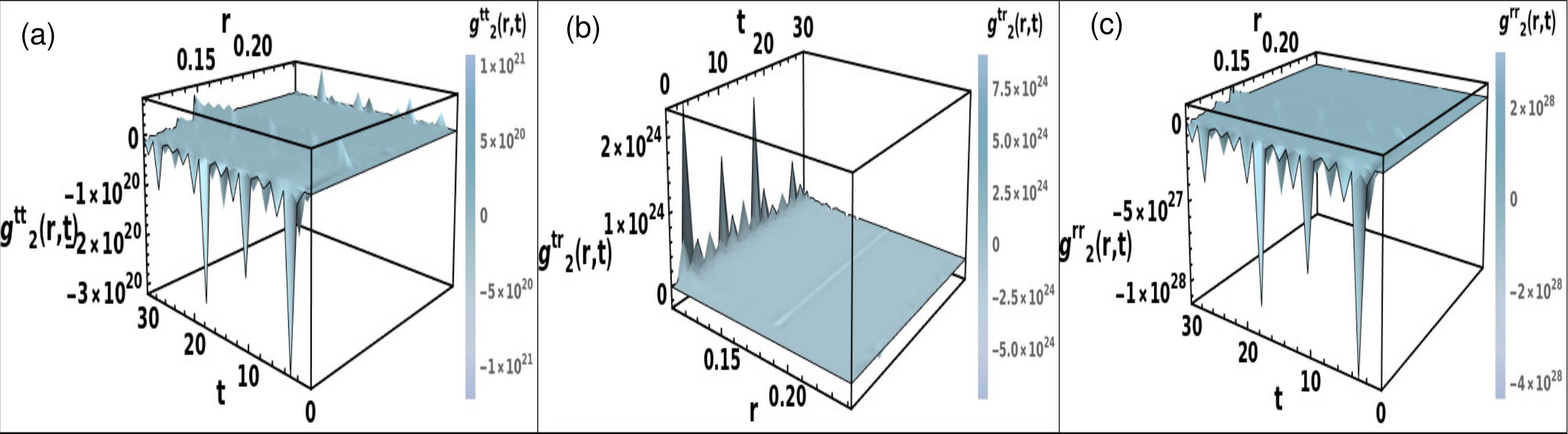}
        \begin{minipage}{\textwidth}
        \caption{Plots displaying all components of the second order inverse acoustic metric (Eq.~\eqref{eq:g2}), derived from our extended fluid model with low frequency perturbation $\omega_{low}=10^{4}~\rm{rad/s}$. Fig.~$12(a)$ illustrates the spacetime behavior of the second order perturbed inverse acoustic metric for the $tt$ component. Fig.~$12(b)$ depicts the spacetime variation of the second order perturbed inverse acoustic metric in the $tr \equiv rt$ components. Fig.~$12(c)$ shows the spacetime variation of the second order perturbed inverse acoustic metric for the $rr$ component.}\label{f:g2all_l}
        \end{minipage}
        \end{center}
        \end{figure*}
    \end{widetext}

We now turn our attention to solving the second-order perturbation equations that describe the Berry curvature induced mass accretion rate variable and charge density, as given in Eq.~\eqref{eq:first_metric} and Eq.~\eqref{eq:first_rho}, respectively, under the assumption of low-frequency perturbations. To compute $F_2(r,t)$, we employ numerical solutions derived from the first order low frequency mass accretion rate $F_1(r,t)$, the first order low frequency inverse acoustic metric components $g^{\mu\nu}_{(1)}$, and the stationary background inverse acoustic metric components $g^{\mu\nu}_{(0)}$. This methodology ensures consistency with the first-order perturbation solution while maintaining the initial condition of exponential temporal decay. Likewise, for the second order density perturbation $\rho_2(r,t)$, we depend solely on the solutions of the second order perturbed mass accretion rate $F_2(r,t)$ to carry out numerical integration. The obtained solutions for $F_2(r,t)$ and $\rho_2(r,t)$ are depicted in Figs.~\ref{f:F2rho2_l}(a) and \ref{f:F2rho2_l}(b), respectively.

By analysing the second order perturbations in the mass accretion rate and density, as well as their first order counterparts (illustrated in Fig.~\ref{f:F1rho1_l}), it is clear that the second order solutions provide a more detailed representation of the first order perturbations, particularly in terms of their amplitudes. A significant observation is that the perturbation amplifies near the acoustic horizon during the initial stages, with the most prominent negative fluctuation in the mass accretion rate occurring around $r\approx 0.1829139~\rm{m}$ in the subsonic region at early times ($t\approx 0~\rm{s}$). 

    \newpage

Apart from this, the oscillatory modes in both subsonic and supersonic flow regions diminish entirely over time.

On the other hand, the density perturbation at low frequency shows slightly stronger growth at this order, as evident from the comparison between Fig.~\ref{f:F2rho2_l}(b) and Fig.~\ref{f:F1rho1_l}(b). Given that the mass accretion rate fluctuation $F_2(r,t)$ reaches its most negative fluctuation value in the subsonic region, the associated density solution $\rho_2(r,t)$ includes small corrections that first emerge in the subsonic zone at this stage. Similar to the behavior of $F_2(r,t)$, the second order density fluctuation grows more significant near the horizon region $(0.1550683~\rm{m} \leq r \leq 0.1859701~\rm{m})$. Additionally, two consecutive second order density fluctuations—observed in their negative regimes—appear at the initial time $t = 0~\rm{s}$ within the supersonic region ($r \approx 0.1550683~\rm{m}$) and the subsonic region ($r \approx 0.1727265~\rm{m}$), respectively.

    \newpage

The inverse metric components at low frequency, calculated up to second order perturbation, are derived by solving Eq.~\ref{eq:g2}. The corresponding results are displayed in Fig.~\ref{f:g2all_l}. A comparison with the first order solutions presented in Fig.~\ref{f:g1all_l} reveals that the observed patterns intensify at this higher order, particularly in terms of their amplitudes.

Fig.~\ref{f:g2all_l} clearly shows that in our extended fluid model system, the second order low frequency perturbations affecting all inverse acoustic metric components are primarily observed in the supersonic flow region close to the inner accretor boundary radius. Unlike the high frequency perturbation solutions for $g^{tt}_{(2)}(r,t)$ and $g^{rr}_{(2)}(r,t)$ at this order (shown in Figs.~\ref{f:g2all_h}(a) and \ref{f:g1all_h}(c), respectively), the perturbations in this case display negative corrections. In particular, for the components $g^{tt}_{(2)}(r,t)$ and $g^{rr}_{(2)}(r,t)$ (presented in Figs.~\ref{f:g2all_l}(a) and \ref{f:g2all_l}(c), respectively), the strongest negative amplitude occurs near time $t \approx 3~\rm{s}$. On the other hand, in contrast to the high frequency perturbation of $g^{tr}_{(2)}(r,t)$, its low frequency counterpart attains its maximum positive value at the same time, $t \approx 3~\rm{s}$.\\
From these findings, we can further deduce that the consistent negative corrections near the inner accretor boundary at $r \approx 0.1068478~\rm{m}$ indicate the existence of a receding acoustic horizon in our studied topological material system when considering second order low frequency perturbations, similar to the behavior observed with first order low frequency perturbations, as previously discussed.  
     
Based on our thorough analysis of the first and second order perturbative solutions describing the Berry curvature induced accretion rate, density, and inverse acoustic metric components in the context of low frequency damped perturbations, we identify a notable phenomenon: a fluctuating acoustic horizon that gradually recedes in size. By extending our investigation to higher order perturbations, we discovered a significant trend suggesting this behavior continues at elevated perturbation orders, at least within the low frequency regime. This leads us to infer that the acoustic horizon's recession likely persists in higher order perturbations as well.\\
Furthermore, it is crucial to highlight that our examination of the first and second order perturbed solutions for the Berry curvature dependent mass accretion rate, charge density, and inverse acoustic metric components within our generalized fluid framework considering both high and low frequency perturbations—reveals a particularly important suppression effect. This effect, emerging from the transition between $\epsilon$ and $\epsilon^2$ terms, becomes especially noteworthy in our study.

To summarize, we find that the position of the acoustic horizon significantly influences both the first and second order perturbed solutions for the Berry curvature-dependent mass accretion rate, density, and inverse acoustic metric components within our extended fluid model. Hence, the topological materials investigated in this work could serve as an experimental tool for exploring such analog dynamical spacetime phenomena.

\section{Discussion}
We have established a fully nonlinear analog gravity framework in topological materials, incorporating Berry curvature into the hydrodynamic flow of charge carriers. By formulating a perturbation scheme with the mass accretion rate as the central variable, we derived a nonlinear wave equation and corresponding time dependent acoustic metric that encodes all nonlinear effects. This approach captures evolving geometric features — including oscillatory and receding horizons, and frequency dependent causal dynamics — beyond the reach of conventional, linearized analog models.

Numerical simulations in graphene reveal the formation and temporal evolution of analog horizons, and the emergence of horizon relaxation and new dynamical regimes. We further demonstrate that the analog Hawking temperature is experimentally accessible, reaching tens of micro-kelvin for realistic parameters. 

This framework significantly advances analog gravity in condensed matter, positioning topological materials as promising platforms to probe nonlinear spacetime dynamics, horizon phenomena, and analog Hawking radiation in laboratory settings.

\newpage
\section*{Acknowledgment}
S. Das gratefully acknowledges the financial support provided by the USTC Fellowship Level A--CAS-ANSO Scholarship 2024 (formerly the ANSO Scholarship for Young Talents) for doctoral studies. Furthermore, S. Das expresses appreciation to Yi-Fu Cai for providing the ``LINDA \& JUDY" computational supports at the Particle Cosmology group (COSPA) at USTC. Generous computing resources were provided by the Sulis HPC service (EP/T022108/1), ARCHER2 UK National Computing Service, which was granted via HPC-CONEXS, the UK High-End Computing Consortium (EPSRC grant no. EP/X035514/1). Numerical computations were performed using the LINDA $\&$ JUDY computing facilities at COSPA, USTC.\\\\\\
  
%\addcontentsline{toc}{section}{References}
\bibliographystyle{apsrev4-1}
\bibliography{main}

\end{document}